\documentclass[journal]{IEEEtran}

\usepackage{amsmath,amssymb}
\usepackage{multicol}
\usepackage{graphicx}
\usepackage{amsmath}
\usepackage{caption}
\usepackage{algorithm}
\usepackage{algpseudocode} 
\usepackage{placeins}
\usepackage{comment}
\usepackage{float}
\usepackage{afterpage}
\usepackage{stfloats}

\usepackage{color}

\usepackage{xcolor}

\title{Polynomial Closed Form Model for Ultra-Wideband Transmission Systems} 

\author{P.~Poggiolini, \IEEEmembership{Fellow, IEEE},
        Y.~Jiang, Y.~Gao,
        and F.~Forghieri, \IEEEmembership{Fellow, IEEE}
\thanks{P. Poggiolini and Y. Jiang are with the Dipartimento di Elettronica e Telecomunicazioni (DET), Politecnico di Torino, 10129, Torino, Italy. Y. Gao is with the Dipartimento di Elettronica e Telecomunicazioni (DET), Politecnico di Torino, 10129, Torino, Italy, and the school of electronic and information engineering, Soochow University, Suzhou, China. F. Forghieri is with CISCO
Photonics, 20871 Vimercate (MB), Italy.} 

\thanks{This work was partially supported by Cisco Systems through the ``BOOST'' Sponsored Research Agreement (SRA) and by the PhotoNext Center of Politecnico di Torino. (Corresponding author: Yanchao Jiang, yanchao.jiang@polito.it. ORCID: https://orcid.org/0009-0009-5082-1257.)}}
\date{August 2025}

\begin{document}

\maketitle
\begin{abstract}
Ultrafast and accurate physical layer models are essential for designing, optimizing and managing ultra-wideband optical transmission systems. We present a closed-form GN/EGN model, named Polynomial Closed-Form Model (PCFM), improving reliability, accuracy, and generality. The key to deriving PCFM is expressing the spatial power profile of each channel along a span as a polynomial. Then, under reasonable approximations, the integral calculation can be carried out analytically, for any chosen degree of the polynomial. We present a full detailed derivation of the model. We then validate it vs. the numerically integrated GN-model in a challenging multiband (C+L+S) scenario, including Raman amplification and inter-channel Raman scattering. We then show that the approach works well also in the special case of the presence of multiple lumped loss along the fiber. Overall, the approach shows very good accuracy and broad applicability. A software implementing the model, fully reconfigurable to any type of system layout, is available for download under the Creative Commons 4.0 License.
\end{abstract}
    
  
\section{Introduction}
\label{sec:Introduction}
Ultra-wideband (UWB) optical transmission systems (also called `multiband') serve as a promising technology to meet the ever-increasing capacity demands of modern communication networks. Systems using C+L bands are being widely deployed. Research is exploring adding further bands, such as S and E, and others \cite{2024_ECOC_Puttnam}-\cite{2025_OFC_Shimizu}. However, UWB systems are affected by very significant inter-channel Raman scattering (ISRS) which together with the deterioration of most propagation parameters towards higher optical frequencies, makes the design of UWB systems challenging. Careful optimization of launch power and other system parameters is then required. In addition, UWB systems greatly benefit from backward Raman amplification, but this in turn requires the optimization of the number, frequency and power of the pumps \cite{2024_PTL_Jiang}, \cite{2025_OFC_Sarkis}.

Optimization is carried out by means of iterative algorithms that require ultra-fast system performance assessment. Non-linear-interference (NLI) estimation is a key component of performance assessment and typically relies on GN/EGN-type models. Since solving the related integrals numerically is not a practicable option due to the large computational effort, these models are implemented as closed-form approximate formulas (CFMs). Over the decade, mainly two groups have pursued the derivation of CFMs, one from University College London (UCL, see \cite{2023_JLT_Buglia}, \cite{2024_JLT_Buglia}), and one from Politecnico di Torino (PoliTo) in collaboration with CISCO (see \cite{2022_ECOC_Poggiolini}, \cite{2024_ECOC_Jiang}). Recently, other groups have started looking into similar investigations too  \cite{2022_JLT_DAmico}-\cite{2023_JLT_Lasagni}. 

This paper reports on a new approach at handling the core integrals of the GN model to obtain a CFM. It consists of expressing the spatial power-profile (SPP) of each channel along a span as a polynomial. Once this is done, some of the core GN model integrals admit closed-form near-exact solutions, involving less drastic approximations than previous CFMs needed to take the SPP into account, especially in the presence of Raman amplification and ISRS. We call the newly derived result `PCFM', where `P' stands for `polynomial'. 

The PCFM can model UWB systems with ISRS, forward and backward Raman amplification, short spans, lumped loss and other possible SPPs perturbations. It enjoys better accuracy and reliability than previous CFMs, because it does not hinge on either neglecting intra-span NLI coherence or resorting to infinite series expansions, or other problematic approximations. The coherent beat of NLI noise along a single span, especially in the presence of Raman amplification, is fully captured, as opposed for instance to our previous most advanced model, CFM6 \cite{2024_ECOC_Jiang}.

In addition, PCFM makes it possible to remove some of the other approximations that CFMs generally make. In particular, the PCFM approach potentially allows to retain the Multi-Channel Interference (MCI) contribution to NLI, making it possible to handle low-dispersion and low-symbol-rate scenarios where the MCI `islands' in the integration domain of the GN and EGN model \cite{2014_OE_Carena} core integrals need to be considered. In principle, it can also allow to remove other approximations that are typically made by CFMs, such as that channel spectra are rectangular and that the power spectral density of NLI is flat over each channel. However, these extensions mentioned in this paragraph are left for a specifically devoted forthcoming submission and will not be dealt with here.

This paper is a follow-up to the OFC 2025 invited paper \cite{2025_OFC_Poggiolini}. Compared to \cite{2025_OFC_Poggiolini}, we provide a full and detailed derivation of the PCFM. Starting with the GNRF (GN-model reference formula), we work out the complete set of integrals for all types of NLI contributions, which can be solved in closed form using the SPP polynomial representation. In particular, the Self-Channel Interference (SCI) and  Cross-Channel Interference (XCI) contributions are explicitly expressed in closed form. We discuss in depth the polynomial fitting of the channel SPPs. We expand on the testing of the model, including the case of multiple lumped losses in a 100~km fiber with significant ISRS and backward Raman amplification. 

This paper is organized as follows. In Sect.~\ref{sec:Premises}, we provide the complete theoretical background involved in the derivation of the PCFM. In Sect.~\ref{sec:PCFM_derivation}, we present the PCFM derivation in detail, focusing on the use of a polynomial to express the SPP of each channel. In Sect.~\ref{sec:PCFM_applications}, we show accuracy tests of SPP representation and of NLI estimation in UWB systems. We provide at the end some hints on what directions the development of the approach is taking and the link to a software implementation which is available for download. Comments and conclusions follow. 

\section{Premises}
\label{sec:Premises}
The NLI noise is produced by the Wavelength Division Multiplexing (WDM) signal in each span. We assume the approximation of \textit{incoherent NLI accumulation}, that is, the NLI produced in each span sums up in power at the end of the link. This approximation can be removed a posteriori, similar to what was done in \cite{2020_JLT_Zefreh}, but it is necessary in this phase. So, the power spectral density (PSD) of NLI at the end of the link, $G_{\text{NLI}}^{\text{end}}\left( f \right)$ is:
\begin{equation}
    \begin{aligned}
    G_{\text{NLI}}^{\text{end}}\left( f \right)\approx \sum\limits_{{{n}_{s}}=1}^{{{N}_{s}}}{G_{\text{NLI}}^{({{n}_{s}}),\text{end}}\left( f \right)}
    \end{aligned} 
    \label{eq:GNLI_IGN}
\end{equation}
where ${N}_{s}$  is the number of spans in the link and the `approximately equal' symbol is used here to point out that incoherent NLI accumulation is an approximation. However, henceforth we will use an `equal-to' symbol for ease of readability. Note that $G_{\text{NLI}}^{(n_s),\text{end}}(f)$ represents the NLI noise produced within the $n_s$-th span, then propagated linearly till the end of the link. 

\begin{figure}
    \centering
    \includegraphics[width=1\linewidth]{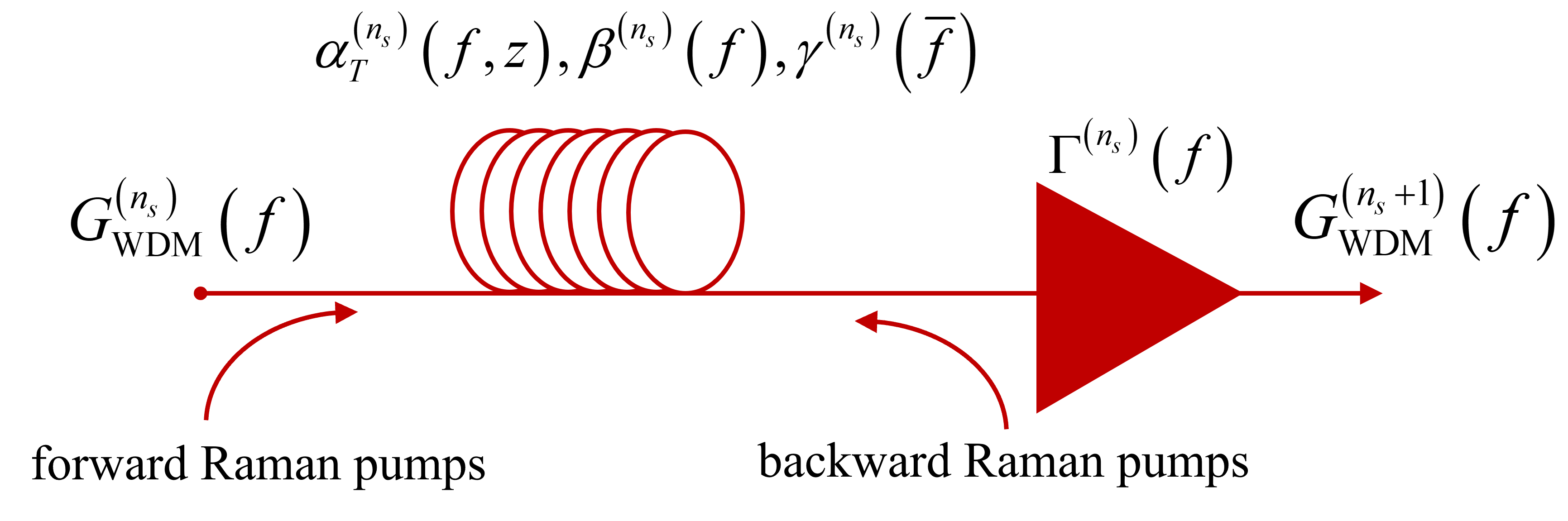}
    \caption{Pictorial representation of a generic span. The symbols appearing in it are explained in the text. }
    \label{fig:generic_span}
\end{figure}

For an illustration of the definition of `span', please see Fig.~\ref{fig:generic_span}. A span is comprised first of a stretch of fiber, which we assume to be \textit{of the same type within the span}. It may of course differ from span to span. The span fiber is characterized through its dispersion $\beta^{(n_s)}(f)$, attenuation $\alpha^{(n_s)}(f)$ and non-linearity coefficient $\gamma^{(n_s)}({\bar{f} })$. Note that $\gamma^{(n_s)}$ depends on multiple frequencies, as it is discussed later, hence the vector symbol above $f$. Both forward and backward-pumped optional Raman amplification may be present in the span, as indicated by the pump injection in Fig.~\ref{fig:generic_span}.

We then make a distinction between fiber attenuation $\alpha^{(n_s)}(f)$, which is assumed $z$-independent, and \textit{total} distributed span attenuation/gain, which also includes possible lumped loss due to splicing and connectors along the span length, as well as forward/backward pumped Raman amplification and ISRS, which is $z$-dependent: $\alpha_{T}^{(n_s)}(f,z)$.

The span then includes a lumped element at the end, shown as a red triangle. Such lumped-element can account for the following: a Doped-Fiber-Amplifier (DFA), a linear filtering element as well as lumped loss before and after amplification. Overall, the lumped-element is modeled using a frequency-dependent lumped power gain/loss $\Gamma^{(n_s)}(f)$. 

At the input of the span, there is the WDM overall signal PSD $G_{\text{WDM}}^{(n_s)}(f)$.  The output of the $n_s\text{-th}$ span is the input of the $(n_s+1)\text{-th}$ span $G_{\text{WDM}}^{(n_s +1)}(f)$. 

We call the transfer function accounting for all linear elements and effects from the input of the $n_s\text{-th}$ span to the output of the ${{m}_{s}}\text{-th}$ span as  ${\left| \text{H}(f;n_s,m_s) \right|}^2$. We can then re-write Eq.~(\ref{eq:GNLI_IGN}) as follows:

\begin{equation}
    \begin{aligned}
    G_{\text{NLI}}^{\text{end}}\left( f \right)=\sum\limits_{{{n}_{s}}=1}^{{{N}_{s}}}{G_{\text{NLI}}^{({{n}_{s}})}\left( f \right)}\cdot {{\left| \text{H}\left( f;{{n}_{s}}+1,{{N}_{s}} \right) \right|}^{2}}
    \end{aligned}
    \label{eq:GNLI_IGN_v2}
\end{equation}
where $G_{\text{NLI}}^{(n_s)}(f)$ is the NLI noise produced in the  $n_s\text{-th}$ span, as it shows up \textit{at the end} of the $n_s\text{-th}$ span, i.e., after the lumped element. 

Note that the last term in the summation has index ${{n}_{s}}={{N}_{s}}$, which generates a transfer function: ${{\left| \text{H}\left( f;{{N}_{s}}+1,{{N}_{s}} \right) \right|}^{2}}$. The fact that the first index is larger than the second may appear to generate an inconsistency. We could easily remove the problem by simply imposing ${{\left| \text{H}\left( f;{{N}_{s}}+1,{{N}_{s}} \right) \right|}^{2}}=1$. There is however a justification for doing so. The writing ${{\left| \text{H}\left( f;{{N}_{s}}+1,{{N}_{s}} \right) \right|}^{2}}$ means that we are looking at the transfer function between two points that coincide: the input of the $\left( {{N}_{s}}+1 \right)$-th span and the output of the ${{N}_{s}}\text{-th}$ span are the same location. Since they spatially coincide, it makes sense that ${{\left| \text{H}\left( f;{{N}_{s}}+1,{{N}_{s}} \right) \right|}^{2}}=1$ .

\subsection{The linear transfer function}\label{subsect:linear_propagation}
We first focus on expressing ${{\left| \text{H}\left( f;n_s,m_s \right) \right|}^{2}}$ in a general closed form. 
To express it, we first point out that:
\begin{equation}
    \begin{aligned}
    {{\left| \text{H}\left( f;{{n}_{s}},{{m}_{s}} \right) \right|}^{2}}=\prod\limits_{n_l={{n}_{s}}}^{{{m}_{s}}}{{{\left| \text{H}\left( f;n_l,n_l \right) \right|}^{2}}}
    \end{aligned}
    \label{eq:transfer_function}
\end{equation}
So, we can concentrate on the simpler problem of finding ${{\left| \text{H}\left( f;n_l,n_l \right) \right|}^{2}}$, which is the square modulus of the linear transfer function from input to output of the $n_l$-th span. We have:
\begin{equation}
    \begin{aligned}
    \text{H}\left( f;n_l,n_l \right)= \sqrt{{{\Gamma }^{(n_l)}}\left( f \right)} \cdot {{e}^{\int_{0}^{L_{s}^{\left( n_l \right)}}{{{\kappa }^{(n_l)}}\left( f,z \right)\,dz}}}
    \end{aligned}
    \label{eq:transfer_function_v2}
\end{equation}
where ${{\kappa }^{(n_l)}}\left( f,z \right)$ is the complex propagation constant defined as:
\begin{equation}
    \begin{aligned}
    {{\kappa }^{(n_l)}}\left( f,z \right)=-j{{\beta }^{(n_l)}}\left( f \right)-{{\alpha_T }^{(n_l)}}\left( f,z \right)
    \end{aligned}
    \label{eq:propagation_constant}
\end{equation}
where, for the $n_l$-th span, ${{\beta }^{(n_l)}}\left( f \right)$ is the propagation constant at frequency $f$ and ${{\alpha_T }^{(n_l)}}\left( f,z \right)$ is the fiber attenuation, at frequency $f$ and location $z$ in the span. Note that in Eq.~(\ref{eq:transfer_function_v2}) it is assumed that the spatial variable $z$ is re-initialized to zero at the start of each span, that is, it is local to the span. It runs from 0 to the $n_l$-th span length $L_{s}^{\left( n_l \right)}$. 

Also, we remind the reader that distributed gain due to ISRS or Raman amplification is incorporated within ${{\alpha_T }^{(n_l)}}\left( f,z \right)$. This means that there may be stretches of fiber whose ${{\alpha_T }^{(n_l)}}\left( f,z \right)<0.$ 

Then Eq.~(\ref{eq:transfer_function_v2}) simplifies to:
\begin{equation}
    \begin{aligned}
    \text{H}(f; n_l, n_l) 
    &= \sqrt{\Gamma^{(n_l)}(f)} \cdot
        e^{-j \beta^{(n_l)}(f) L_s^{(n_l)}} \,\\
    &  \cdot e^{-\int_{0}^{L_s^{(n_l)}} \alpha_T^{(n_l)}(f,z) \, dz}
    \end{aligned}
    \label{eq:transfer_function_v3}
\end{equation}
Eq.~(\ref{eq:transfer_function}) then reads:
\begin{equation}
    \begin{aligned}
    {{\left| \text{H}\left( f;{{n}_{s}},{{m}_{s}} \right) \right|}^{2}}=\prod\limits_{n_l={{n}_{s}}}^{{{m}_{s}}}{{{\Gamma }^{(n_l)}}\left( f \right){{e}^{-2\int_{0}^{L_{s}^{\left( n_l \right)}}{{{\alpha_T }^{(n_l)}}\left( f,z \right)dz}}}}
    \end{aligned}
    \label{eq:transfer_function_v4}
\end{equation}
and note the important aspect that, due to the absolute value squared on the left-hand side, dispersion disappears completely from Eq.~(\ref{eq:transfer_function_v4}). As a consequence, it also disappears from Eq.~(\ref{eq:GNLI_IGN_v2}) which becomes:
\begin{equation}
    \begin{aligned}
    G_{\text{NLI}}^{\text{end}}\left( f \right) & = \sum\limits_{{{n}_{s}}=1}^{{{N}_{s}}}{G_{\text{NLI}}^{({{n}_{s}})}\left( f \right)} \\
   & \cdot \prod\limits_{n_l={{n}_{s}}+1}^{{{N}_{s}}}{{{\Gamma }^{(n_l)}}\left( f \right){{e}^{-2\int_{0}^{L_{s}^{\left( n_l \right)}}{{{\alpha_T }^{(n_l)}}\left( f,z \right)dz}}}}
    \end{aligned}
    \label{eq:GNLI_IGN_v3}
\end{equation}

This equation embodies our fundamental premises, including the incoherent NLI accumulation assumption. If all spans were transparent, i.e., with gain exactly compensating for loss, at all frequencies, the linear propagation factors would all be 1 and NLI at the receiver would simply be:
\begin{equation}
G_{\text{NLI}}^{\text{end}}\left( f \right) = \sum\limits_{{{n}_{s}}=1}^{{{N}_{s}}}{G_{\text{NLI}}^{({{n}_{s}})}\left( f \right)}
\end{equation}

Of course, if transparency is not verified, Eq.~(\ref{eq:GNLI_IGN_v3}) accounts for the general case.

\subsection{NLI PSD}\label{subsect:NLI_PSD} 
According to the assumptions and derivations of the incoherent GN model \cite{2020_chapter_Bononi}, Eq.~(\ref{eq:GNLI_ns}) provides the NLI PSD produced by the generic $n_s\text{-th}$ span, $G_{\text{NLI}}^{({{n}_{s}})}\left( f \right)$. 

The integral contains the factors $(\gamma^{(n_{s})})^2$ and $\left| \rho^{(n_{s})}\right|^2 $ which will be discussed later. We first focus instead on the role of the WDM signal $G_{\text{WDM}}^{(n_s)}(f)$ at the input of the $n_s\text{-th}$ span, which appears three times in the integrand function of Eq.~(\ref{eq:GNLI_ns}). An example of a possible $G_{\text{WDM}}^{(n_s)}(f)$ is shown in Fig.~\ref{fig:WDM_spectrum}. Note that the superscript $n_{s}$ is present because we assume that in each span there can be a different set of WDM channels, that is, a different $G_{\text{WDM}}^{(n_s)}(f)$. We also assume that the channel under test (CUT) is present in each span, at the same frequency. All other channels can change, in number, frequency and bandwidth. Specifically, the number of WDM channels in the ${{n}_{s}}\text{-th}$ span is $N_{\text{ch}}^{({{n}_{s}})}$. The set of center frequencies and the set of channel bandwidths in the ${{n}_{s}}\text{-th}$ span are:
\begin{equation*}
    \left\{ f_{{n_{\text{ch}}}}^{(n_{s})} \right\}_{n_{\text{ch}}=1}^{N_{\text{ch}}^{(n_{s})}}, \quad
    \left\{ B_{{n_{\text{ch}}}}^{(n_{s})} \right\}_{n_{\text{ch}}=1}^{N_{\text{ch}}^{(n_{s})}}
\end{equation*}

\begin{figure*} 
\begin{equation}
    \begin{aligned}
      G_{\text{NLI}}^{({{n}_{s}})}\left( f \right) & =\frac{16}{27}{{\Gamma }^{({{n}_{s}})}}\left( f \right)\int_{-\infty }^{\infty }{\int_{-\infty }^{\infty }{G_{\text{WDM}}^{({{n}_{s}})}\left( {{f}_{1}} \right)G_{\text{WDM}}^{({{n}_{s}})}\left( {{f}_{2}} \right) G_{\text{WDM}}^{({{n}_{s}})}\left( {{f}_{1}}+{{f}_{2}}-f \right)}} \\
      & \cdot {{\left( {{\gamma }^{({{n}_{s}})}}\left( {{f}_{1}},{{f}_{2}},f \right) \right)}^{2}} \cdot {{\left| \rho ^{({{n}_{s}})} \left( {{f}_{1}},{{f}_{2}},f \right) \right|}^{2}}d{{f}_{1}}d{{f}_{2}} \\ 
    \\
    {\left| \rho ^{({{n}_{s}})} \left( {{f}_{1}},{{f}_{2}},f \right) \right|}^{2} &= {{\left| {{e}^{-\int_{0}^{L_{s}^{\left( {{n}_{s}} \right)}}{{{\alpha_T }^{({{n}_{s}})}}\left( f,z \right)dz}}}\int_{0}^{L_{s}^{\left( {{n}_{s}} \right)}}{{{e}^{\int_{0}^{z}{\Delta {{\kappa }^{({{n}_{s}})}}\left( {{f}_{1}},{{f}_{2}},f,z' \right) d{z}'}}}}dz \right|}^{2}}\\ 
\end{aligned}
\label{eq:GNLI_ns}
\end{equation}
\end{figure*}

\begin{figure}
    \centering\hspace{-0.7cm}
    \includegraphics[width=0.9\linewidth]{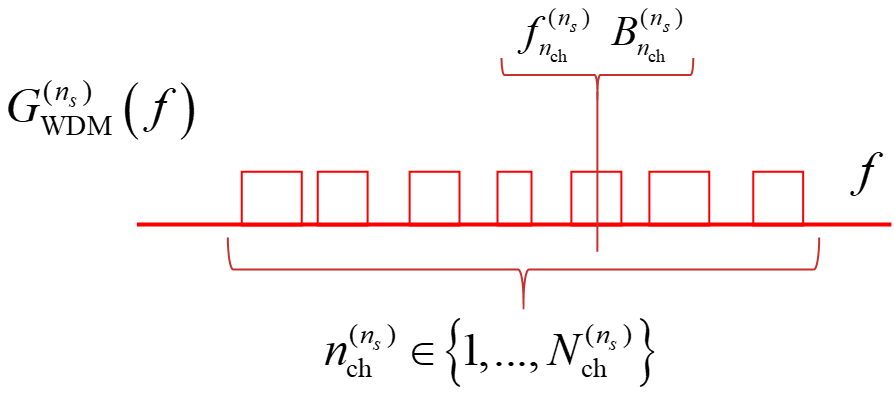}\vspace{0.5cm}
    \caption{Pictorial representation of an example of a WDM spectrum $G_{\text{WDM}}^{({{n}_{s}})}\left( f \right)$. Symbols are explained in the text.}
    \label{fig:WDM_spectrum}
\end{figure}

We explicitly write the CUT channel index in the ${{n}_{s}}\text{-th}$ span as $n_{\text{CUT}}^{({{n}_{s}})}$. Of course, $n_{\text{CUT}}^{({{n}_{s}})}$ is one of the channel indices of that span, that is $n_{\text{CUT}}^{({{n}_{s}})}\in \left\{ 1,\ldots ,N_{\text{ch}}^{({{n}_{s}})} \right\}$. The frequency and bandwidth of the CUT in the ${{n}_{s}}\text{-th}$ span should then be written $f_{n_{\text{CUT}}^{({{n}_{s}})}}^{({{n}_{s}})}$, $B_{n_{\text{CUT}}^{({{n}_{s}})}}^{({{n}_{s}})}$. 
However, the frequency and bandwidth of the CUT do not change span-by-span, so for ease of notation we will write them as: ${{f}_{\text{CUT}}}$, ${{B}_{\text{CUT}}}$, dropping all indices.

We then approximate the channel spectra as rectangles, whose bandwidth is $B_{{{n}_{\text{ch}}}}^{({{n}_{s}})}$. As a result, we can write the WDM comb at the input of the ${{n}_{s}}\text{-th}$ span as:
\begin{equation}
    \begin{aligned}
    G_{\text{WDM}}^{({{n}_{s}})}\left( f \right)=\sum\limits_{{{n}_{\text{ch}}}=1}^{N_{\text{ch}}^{({{n}_{s}})}}{G_{\text{WDM,}{{n}_{\text{ch}}}}^{({{n}_{s}})}\cdot {{\Pi }_{B_{{{n}_{\text{ch}}}}^{({{n}_{s}})}}}\left( f-f_{{{n}_{\text{ch}}}}^{({{n}_{s}})} \right)}
    \end{aligned}
    \label{eq:WDM_spectrum}
\end{equation}
where ${{\Pi }_{B}}\left( f \right)$ is a Heaviside `pi' function (i.e., a rectangle) of bandwidth $B$, defined as:
\begin{equation}
    \Pi_{B}(f) = 
    \begin{cases}
        1, & f \in \left[-\frac{B}{2}, \frac{B}{2} \right] \\
        0, & \text{otherwise}
    \end{cases}
\end{equation}
and $G_{\text{WDM},n_{\text{ch}}}^{({{n}_{s}})}$ is the value of the channel PSD, setting the height of the rectangular spectrum.

Regarding the approximation of a channel spectrum with a rectangle in the context of the GN-model, this topic has been discussed in depth in the past, and possibly for the first time in \cite{2012_JLT_Poggiolini}. As time goes by, this approximation seems to become increasingly immaterial, as modern transceivers operate at very small roll-offs, typically on the order of 0.1 or less. Throughout this paper we assume rectangular spectra.

Introducing Eq.~(\ref{eq:WDM_spectrum}) into Eq.~(\ref{eq:GNLI_ns}) we get:
\begin{equation}
\begin{aligned}
  & G_{\text{NLI}}^{(n_s)}(f) = \frac{16}{27} \Gamma^{(n_s)}(f)\\
  &\sum_{m_{\text{ch}}=1}^{N_{\text{ch}}^{(n_s)}}
  \sum_{k_{\text{ch}}=1}^{N_{\text{ch}}^{(n_s)}}
  \sum_{n_{\text{ch}}=1}^{N_{\text{ch}}^{(n_s)}}
   G_{\text{WDM}, m_{\text{ch}}}^{(n_s)} G_{\text{WDM}, k_{\text{ch}}}^{(n_s)}  G_{\text{WDM}, n_{\text{ch}}}^{(n_s)} \\
  & \int_{-\infty}^{\infty} \int_{-\infty}^{\infty}
\Pi_{B_{m_{\text{ch}}}^{(n_s)}}(f_1 - f_{m_{\text{ch}}}^{(n_s)})
  \Pi_{B_{k_{\text{ch}}}^{(n_s)}}(f_2 - f_{k_{\text{ch}}}^{(n_s)}) \\
  &  \cdot
  \Pi_{B_{n_{\text{ch}}}^{(n_s)}}(f_1 + f_2 - f - f_{n_{\text{ch}}}^{(n_s)})  
  \left( \gamma^{(n_s)}(f_1, f_2, f) \right)^2 \\
  & \cdot
  \left| \rho^{(n_s)}(f_1, f_2, f) \right|^2
  \, df_1 df_2
\end{aligned}
\label{eq:GNLI_ns_v2}
\end{equation}

We then concentrate on calculating $G_{\text{NLI}}^{({{n}_{s}})}\left( f \right)$ related to a single channel, specifically the CUT. Note that the CUT can be any of the WDM comb channels, so there is actually no loss of generality in this assumption. 

We also limit our interest to finding the PSD of NLI at the \textit{center frequency} of the CUT, that is $G_{\text{NLI}}^{({{n}_{s}})}\left( {{f}_{\text{CUT}}} \right)$. Given the formalism that we are introducing in this paper, it would be possible to calculate in closed-form the PSD of NLI at frequencies other than the center frequency of the CUT. On the other hand, it has been consistently found (for the first time in \cite{2012_JLT_Poggiolini}) that, while not perfectly, $G_{\text{NLI}}^{({{n}_{s}})}\left( f \right)$ is rather flat over the bandwidth occupied by each channel. As a result, the assumption of `locally white' NLI noise over each channel, with a PSD value equal to the value that it assumes at the center frequency of that channel, is an accepted approximation. We do however mean to investigate the possibility of removing such approximation, in a forthcoming study.

\begin{figure}
    \centering

    \includegraphics[width=0.7\linewidth]{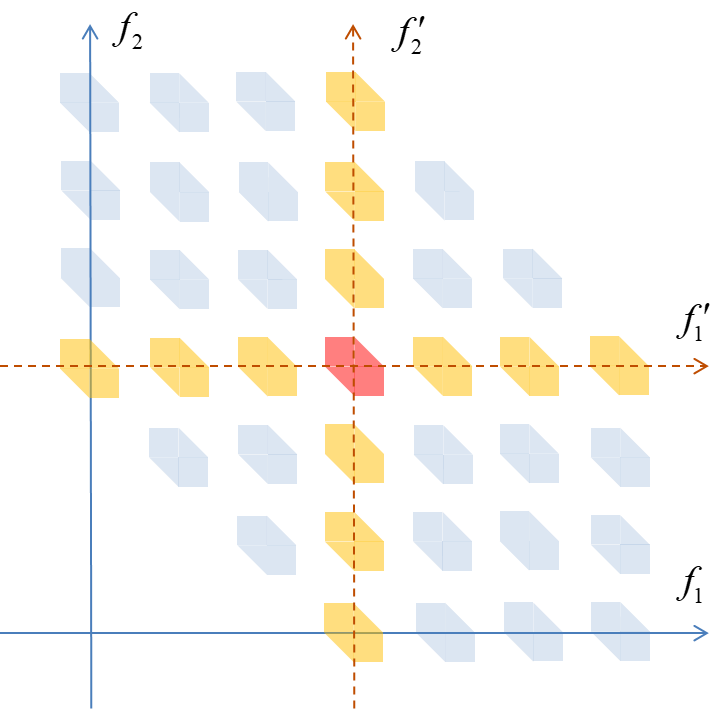}
    \vspace{3mm}

    \includegraphics[width=0.7\linewidth]{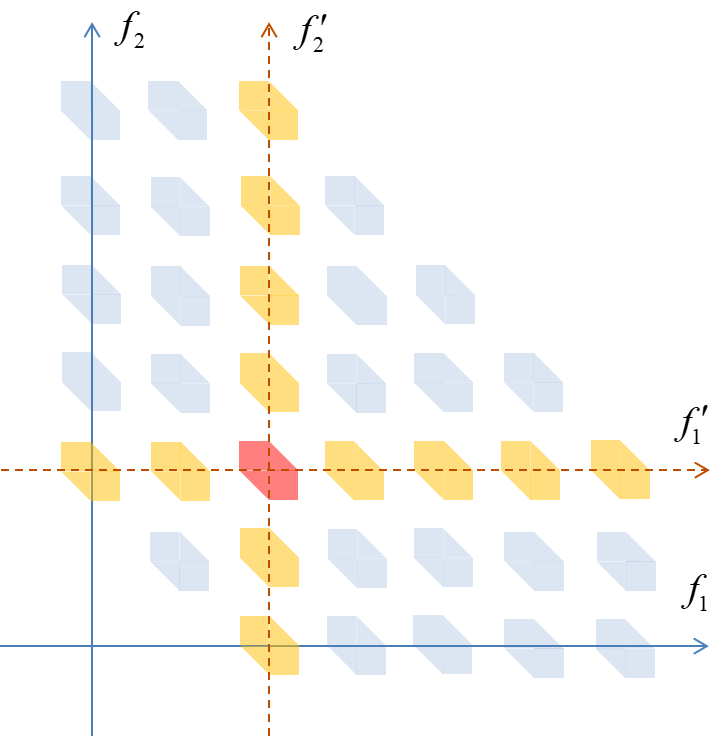}
    \vspace{3mm}
    
    \caption{Integration domain `islands' of Eq.~(\ref{eq:GNLI_ns_v2}), for a 7-channel WDM comb of equally spaced and identical-bandwidth channels. In red the SCI island, in yellow the XCI islands, in light blue the MCI islands. \textbf{Top}: the CUT is the center channel in the WDM comb. \textbf{Bottom}: the CUT is the channel to the left of the center channel in the WDM comb. In both plots, the origin of the frequency axes $(f_{1},f_{2})$ is set to the frequency of the lowest-frequency WDM channel, whereas the origin of the $(f'_{1},f'_{2})$ axes is set to the center frequency of the CUT (see origin shift, Eq.~(\ref{eq:new_axes})). Note that the choice of where to place the origin is irrelevant as to the result of the calculations and is only a matter of convenience.}
    \label{fig:integration_domain}
\end{figure}

Each of the terms of the triple summation then creates a specific integration domain in ${{f}_{1}}$ and ${{f}_{2}}$. Specifically, considering a generic term of the triple summation, the integration domain for ${{f}_{1}}$ and ${{f}_{2}}$ results from the intersection of three conditions:
\begin{equation}
    \begin{aligned}
        f_{1} &\in \left[ f_{m_{\text{ch}}}^{(n_{s})} - \frac{B_{m_{\text{ch}}}^{(n_{s})}}{2},\ 
                        f_{m_{\text{ch}}}^{(n_{s})} + \frac{B_{m_{\text{ch}}}^{(n_{s})}}{2} \right] \\
               &\hspace{-5mm}\text{which is where } \Pi_{B_{m_{\text{ch}}}^{(n_{s})}} \left( f_{1} - f_{m_{\text{ch}}}^{(n_{s})} \right)\ne 0, \\               
        f_{2} &\in \left[ f_{k_{\text{ch}}}^{(n_{s})} - \frac{B_{k_{\text{ch}}}^{(n_{s})}}{2},\ 
                        f_{k_{\text{ch}}}^{(n_{s})} + \frac{B_{k_{\text{ch}}}^{(n_{s})}}{2} \right] \\
               &\hspace{-5mm}\text{which is where } \Pi_{B_{k_{\text{ch}}}^{(n_{s})}} \left( f_{2} - f_{k_{\text{ch}}}^{(n_{s})} \right)\ne0, \\
        f_{1} + f_{2} &\in \left[ f_{\text{CUT}} + f_{n_{\text{ch}}}^{(n_{s})} - \frac{B_{n_{\text{ch}}}^{(n_{s})}}{2},\ 
                                 f_{\text{CUT}} + f_{n_{\text{ch}}}^{(n_{s})} + \frac{B_{n_{\text{ch}}}^{(n_{s})}}{2} \right]\\
               &\hspace{-5mm}\text{which is where } \Pi_{B_{n_{\text{ch}}}^{(n_{s})}} \left( f_{1} + f_{2} - f_{\text{CUT}} - f_{n_{\text{ch}}}^{(n_{s})} \right) \ne 0.
        \end{aligned}
        \label{eq:island_conditions}
\end{equation}
The intersection of the above three conditions creates distinct sub-domains, sometimes called `islands'. Looking for instance at an example of a 7-channel WDM comb, made up of equally-spaced, identical-bandwidth channels, where the CUT is the center channel, then the integration islands that contribute to the NLI PSD at the center of the CUT are those shown in Fig.~\ref{fig:integration_domain}-top. 

Since each island accounts for the NLI which is produced by the `beating' of three specific channels of the WDM comb, We can therefore label each island by means of an identifier $x$ made up of a triple of integers, where each integer is the index of one of the three WDM channels that produce the NLI contribution related to that specific island. Not all triples of channels create an island, though, because the intersection of the conditions Eq.~(\ref{eq:island_conditions}) may turn up empty. The set $\mathcal{X}$ of the triples of channel indices $x$ that do create an islands for a given CUT is formally:
\begin{equation*}
    \begin{aligned}
        x & = \quad  (m_{\text{ch}}^{(n_{s})}, k_{\text{ch}}^{(n_{s})}, n_{\text{ch}}^{(n_{s})}) \in \mathcal{X} \\
        \mathcal{X} & = \quad  \{ (m_{\text{ch}}^{(n_{s})}, k_{\text{ch}}^{(n_{s})}, n_{\text{ch}}^{(n_{s})}) \in \mathbb{N}^3 \mid  \\ 
        &1 \leq m_{\text{ch}}^{(n_{s})}, k_{\text{ch}}^{(n_{s})}, n_{\text{ch}}^{(n_{s})} \leq {N_{\text{ch}}^{(n_{s})}}  \\ 
        &\&\,\,\,f_{\text{CUT}} =   f_{m_{\text{ch}}}^{(n_{s})} + f_{k_{\text{ch}}}^{(n_{s})} - f_{n_{\text{ch}}}^{(n_{s})}   \}   
    \end{aligned}
\end{equation*}

Degeneracy is possible, in the sense that the three indices need not be all different. With reference to the example of Fig.~\ref{fig:integration_domain}, the red island has indeed all three indices identical, corresponding to the channel index of the CUT, $n^{(n_{s})}_{\rm CUT}$. This island accounts for the beating of the CUT with itself, i.e., SCI. The yellow islands have one index that is the CUT and the other two indices identical, spanning all WDM channels other than the CUT. These islands account for the NLI caused by each single channel onto the CUT, called XCI. The light blue islands have all three indices different from that of the CUT and account for MCI.  

The NLI PSD on the CUT in Eq.~(\ref{eq:GNLI_ns_v2}) is calculated by summing up the contribution $G_{\text{NLI,}x}^{(n_s)}\left( f_{\text{CUT}} \right)$ of each individual island $x$, as follows:
\begin{equation}
    G_{\text{NLI}}^{(n_{s})}\left( f_{\text{CUT}} \right) =  \sum_{\substack{x} \in \mathcal{X}} 
    G_{\text{NLI,}x}^{(n_s)}\left( f_{\text{CUT}} \right)
\label{eq:NLI_contribution_for_all_islands_lozenges}
\end{equation}

\begin{figure}
    \centering
    \includegraphics[width=0.7\linewidth]{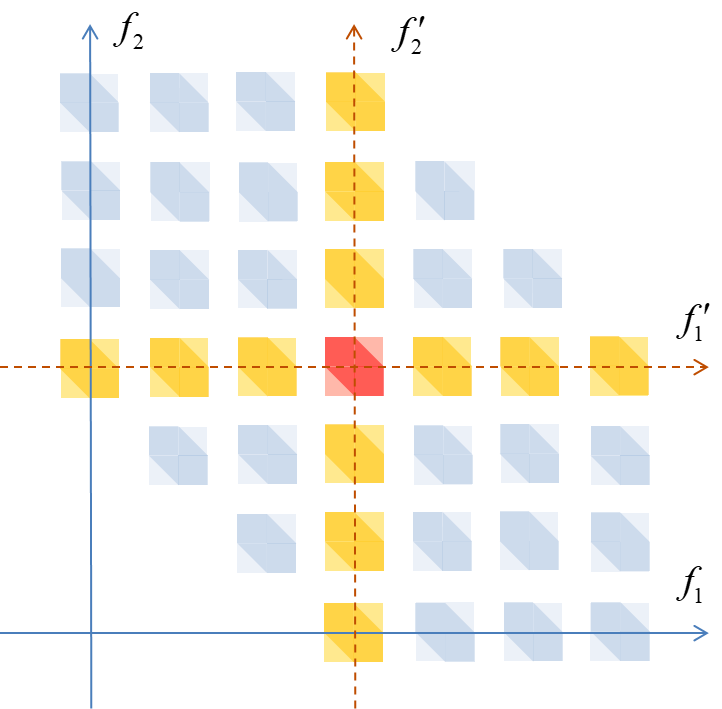}
    \caption{The lighter colored shading shows how the lozenge-shaped integration domains are approximated as squares, with reference to the case shown in the example of Fig.~\ref{fig:integration_domain}-top.}
    \label{fig:rectangles}
\end{figure}

Each contribution therefore requires integrating over a lozenge-shaped island like the ones in Fig.~\ref{fig:integration_domain}. However, this is quite challenging to do analytically. The problem would be much simpler if we could approximate each island with the square or rectangle that inscribes it, such as shown in Fig.~\ref{fig:rectangles}.

It can be shown that this approximation does not cause excessive error, for various reasons. One reason is the interesting aspect that the lozenge-shaped integration islands tend to become rectangle-shaped when the channel spacing $\Delta f$ becomes close to the channel bandwidth $B_{\rm ch}$. Specifically, this manifests itself with the appearance of smaller triangular sub-islands, when $\Delta f < 3/2 \cdot B_{\rm ch}$, which then grow and merge with the lozenge-shaped islands to form a rectangle (\cite{2012_JLT_Poggiolini}, Fig. 20). Given the trend towards ultra-high Baud-rate channels, with tight spacing, this condition is increasingly met.

Besides this, there are also other reasons that justify this approximation, even when the spacing is not so tight, based on the decay features of the integrand function $\rho^{(n_s)} \left(f_1,f_2,f_{\text{CUT}} \right)$, originally pointed out in \cite{2012_JLT_Poggiolini}. We will come back to this aspect later.

Converting the islands into the inscribing rectangles, we get:
\begin{equation}
\begin{aligned}
    & G_{\text{NLI,}x}^{(n_s)}\left( f_{\text{CUT}} \right) = \frac{16}{27} \Gamma^{(n_{s})}\left( f_{\text{CUT}} \right) 
    G_{\text{WDM,}m_{\text{ch}}}^{(n_{s})} G_{\text{WDM,}k_{\text{ch}}}^{(n_{s})}\\
    & \cdot G_{\text{WDM,}n_{\text{ch}}}^{(n_{s})} 
    \int_{f_{k_{\text{ch}}}^{(n_{s})} - B_{k_{\text{ch}}}^{(n_{s})}/2}^{f_{k_{\text{ch}}}^{(n_{s})} + B_{k_{\text{ch}}}^{(n_{s})}/2}
    \int_{f_{m_{\text{ch}}}^{(n_{s})} - B_{m_{\text{ch}}}^{(n_{s})}/2}^{f_{m_{\text{ch}}}^{(n_{s})} + B_{m_{\text{ch}}}^{(n_{s})}/2} \\
    &\left( \gamma^{(n_{s})}(f_{1}, f_{2}, f_{\text{CUT}}) \right)^2
    \left| \rho^{(n_s)}(f_{1}, f_{2}, f_{\text{CUT}}) \right|^2 \, df_{1} df_{2} \\
\end{aligned}
\label{eq:NLI_contribution_for_all_islands}
\end{equation}
We will focus on this integral to derive its closed-form solution for any generic island $x$ in the integration domain.

\section{PCFM derivation}
\label{sec:PCFM_derivation}
Focusing on the integrand in Eq.~(\ref{eq:NLI_contribution_for_all_islands}), we find the fiber non-linearity coefficient $\gamma^{(n_s)}(f_1,f_2,{f_{\text{CUT}})}$. It is indicated as a function of three frequencies but it actually depends on four frequencies. However, in this context, the fourth frequency would be $(f_1 + f_2 - f_{\rm CUT})$. Since it is a function of the other three, we omit to indicate it.

As a reasonable approximation, we assume that $\gamma^{(n_s)}$ is a constant over each island $x$. Then, according to \cite{2010_OE_Santagiustina}, $\gamma^{(n_s)}_{x}$ can be written as: 
\begin{equation}
    \begin{aligned} 
    & \gamma^{(n_s)}_x \overset{\triangle}{=} \gamma^{(n_s)}(f_1,f_2,{f_{\text{CUT}})} \approx \gamma^{(n_s)} (f_{m_{\text{ch}}}^{(n_s)},f_{k_{\text{ch}}}^{(n_s)},{f_{\text{CUT}})} \\
    & = \frac{2\pi f_{\text{CUT}}}{c} \cdot \frac{n_2}{A_{\text{eff}}(f_{\text{CUT}}, f_{m_{\text{ch}}}^{(n_s)},f_{k_{\text{ch}}}^{(n_s)},f_{n_{\text{ch}}}^{(n_s)})} 
    \end{aligned}
    \label{eq:gamma_approximation}
\end{equation}
where $c$ denotes the speed of light in vacuum and $n_2$ is the nonlinear refractive index of the fiber. The quantity ${A_{\text{eff}}(f_{\text{CUT}}, f_{m_{\text{ch}}}^{(n_s)},f_{k_{\text{ch}}}^{(n_s)},f_{n_{\text{ch}}}^{(n_s)})} $ is the cross-effective area of the four frequency components involved in the non-linear Kerr interaction, which can be approximately expressed as \cite{2010_OE_Santagiustina}, \cite{2003_JLT_Rottwitt}:
\begin{equation}
    \begin{aligned} 
    & A_{\text{eff}}(f_{\text{CUT}}, f_{m_{\text{ch}}}^{(n_s)},f_{k_{\text{ch}}}^{(n_s)},f_{n_{\text{ch}}}^{(n_s)}) \\ 
    & \approx \frac{A_{\text{eff}} (f_{\text{CUT}}) + A_{\text{eff}} (f_{m_{\text{ch}}}^{(n_s)}) + A_{\text{eff}} (f_{k_{\text{ch}}}^{(n_s)}) + A_{\text{eff}} (f_{n_{\text{ch}}}^{(n_s)})}{4}
    \end{aligned}
    \label{eq:eff_areas}
\end{equation}
The above formulas show that $\gamma^{(n_s)}$ depends on the type of NLI contribution. For the SCI island, for instance, all four effective areas at the numerator of Eq.~(\ref{eq:eff_areas}) are equal to $A_{\text{eff}}(f_{\text{CUT}})$ and the resulting value of $\gamma^{(n_s)}$ is the one that is generally used in models:
\begin{equation}\gamma^{(n_s)}_{\rm SCI}= \frac{2\pi f_{\text{CUT}}}{c} \cdot \frac{n_2}{A_{\text{eff}}(f_{\text{CUT}})}
\label{eq:gamma_SCI}
\end{equation}
For the other islands, somewhat different values are found, that relate to XCI and MCI. To compute the $A_{\text{eff}}$ terms in Eq.~(\ref{eq:eff_areas}), we used \cite{2022_OE_Shevchenko} Eq.~(4) (derived from \cite{1977_BellTech_Marcuse}). Note that $A_{\text{eff}}$ depends on the type of fiber and hence on the span index $n_s$, but we omitted it to avoid clutter.

Since $\gamma^{(n_s)}$ is now a constant in each island $x$, we can pull it out of the integrals:
\begin{equation}
    \begin{aligned} 
         & G_{\text{NLI,}x}^{(n_s)}\left( f_{\text{CUT}} \right) = \frac{16}{27} \, \Gamma^{(n_s)}\left( f_{\text{CUT}} \right) G_{\text{WDM,}m_{\text{ch}}}^{(n_s)} \\ 
         &\cdot G_{\text{WDM,}k_{\text{ch}}}^{(n_s)} \, G_{\text{WDM,}n_{\text{ch}}}^{(n_s)}
          \left( \gamma^{(n_s)}_x \right)^2 \\
          &\int_{f_{k_{\text{ch}}}^{(n_s)} - B_{k_{\text{ch}}}^{(n_s)}/2}^{f_{k_{\text{ch}}}^{(n_s)} + B_{k_{\text{ch}}}^{(n_s)}/2} 
          \int_{f_{m_{\text{ch}}}^{(n_s)} - B_{m_{\text{ch}}}^{(n_s)}/2}^{f_{m_{\text{ch}}}^{(n_s)} + B_{m_{\text{ch}}}^{(n_s)}/2} 
         \left| \rho^{(n_s)} \left( f_1, f_2, f_{\text{CUT}} \right) \right|^2 \, df_1 \, df_2 \\
   \end{aligned}
  \label{eq:NLI_integral_x}
\end{equation}

We then focus on the link function $\rho^{(n_s)}(f_1, f_2,f_{\text{CUT}})$:
\begin{equation}
    \begin{aligned} 
        \rho^{(n_s)}&(f_1, f_2, f_{\text{CUT}})  = 
        e^{-\int_{0}^{L_s^{(n_s)}} \alpha_T^{(n_s)}(f_{\text{CUT}}, z) \, dz} \\
        & \int_{0}^{L_s^{(n_s)}} 
        e^{\int_{0}^{z} \Delta \kappa^{(n_s)}(f_1, f_2, f_{\text{CUT}}, z') \, dz'} \, dz \\
      \Delta {{\kappa }^{({{n}_{s}})}} & \left( {{f}_{1}},{{f}_{2}},f_{\text{CUT}},z \right)  = 
      -j\Delta {{\beta }^{({{n}_{s}})}}\left( {{f}_{1}},{{f}_{2}},f_{\text{CUT}} \right) \\ &-\Delta {{\alpha_T }^{({{n}_{s}})}}\left( {{f}_{1}},{{f}_{2}},f_{\text{CUT}},z \right) \\
     \Delta {{\beta }^{({{n}_{s}})}}&\left( {{f}_{1}},{{f}_{2}},f_{\text{CUT}} \right) = {{\beta }^{({{n}_{s}})}}\left( {{f}_{1}} \right) +{{\beta }^{({{n}_{s}})}}\left( {{f}_{2}} \right)\\
    &-{{\beta }^{({{n}_{s}})}}\left( f_{\text{CUT}} \right)-{{\beta }^{({{n}_{s}})}}\left( {{f}_{1}}+{{f}_{2}}-f_{\text{CUT}} \right) \\ 
    \Delta {{\alpha_T }^{({{n}_{s}})}}&\left( {{f}_{1}},{{f}_{2}},f_{\text{CUT}},z \right)  ={{\alpha_T }^{({{n}_{s}})}}\left( {{f}_{1}},z \right)+{{\alpha_T }^{({{n}_{s}})}}\left( {{f}_{2}},z \right)\\
   & -{{\alpha_T }^{({{n}_{s}})}}\left( f_{\text{CUT}},z \right)+{{\alpha_T }^{({{n}_{s}})}}\left( {{f}_{1}}+{{f_{2}}-f_{\text{CUT}}},z \right) 
\end{aligned}
\label{eq:link_function}
\end{equation}
In the following we discuss the various quantities appearing in the above formula.

\subsection{Dispersion}
To accurately capture dispersion in UWB systems, we assume that it is expressed through its fourth-order series expansion as: 
\begin{equation}
\begin{aligned}
   &{{\beta }^{({{n}_{s}})}}\left( f \right)  = \beta _{0}^{\left( {{n}_{s}} \right)}+2\pi \beta _{1}^{\left( {{n}_{s}} \right)}\left( f-f_{c}^{({{n}_{s}})} \right)\\
        &+2{{\pi }^{2}}\beta _{2}^{\left( {{n}_{s}} \right)}{{\left( f-f_{c}^{({{n}_{s}})} \right)}^{2}} 
        +\frac{4}{3}{{\pi }^{3}}\beta _{3}^{\left( {{n}_{s}} \right)}{{\left( f-f_{c}^{({{n}_{s}})} \right)}^{3}}\\
        & +\frac{2}{3}{{\pi }^{4}}\beta _{4}^{\left( {{n}_{s}} \right)}{{\left( f-f_{c}^{({{n}_{s}})} \right)}^{4}}+O{{\left( f-f_{c}^{({{n}_{s}})} \right)}^{5}} \\
\end{aligned}  
\label{eq:15_4th-ordered expanded beta}
\end{equation}
where $f_{c}^{({{n}_{s}})}$ is the arbitrary frequency where the expansion is taken in the ${{n}_{s}}\text{-th}$ span. Substituting Eq. (\ref{eq:15_4th-ordered expanded beta}) into the expression of $\Delta {{\beta }^{({{n}_{s}})}}$ in Eq. (\ref{eq:link_function}), we then get the expression:
\begin{equation}
    \begin{aligned}
        &\Delta \beta^{(n_s)}\bigl( f_1, f_2,f_{\text{CUT}} \bigr) = -4\pi^2(f_1-f_{\text{CUT}} )(f_2-f_{\text{CUT}}) \\
        &\cdot \Bigl( \beta_2^{(n_s)} + \pi \beta_3^{(n_s)} \bigl( f_1 + f_2 - 2f_c^{(n_s)} \bigr) + \frac{2}{3} \pi^2 \beta_4^{(n_s)} \\
        &\cdot \Bigl( \bigl( f_1 - f_c^{(n_s)} \bigr)^2 + \bigl( f_1 - f_c^{(n_s)} \bigr) \bigl(f_2 - f_c^{(n_s)} \bigr) \\
        &+ \bigl(f_2 - f_c^{(n_s)} \bigr)^2 \Bigr) \Bigr)
    \end{aligned}
\label{eq:delta_beta}
\end{equation}

For reasons of convenience, we now change the integration variables of Eq.~(\ref{eq:NLI_integral_x}) as follows:
\begin{equation}
    f_1 = f_1' + f_{\text{CUT}}\,\,\,\,\, ,\,\,\,\,\,  f_2 = f_2' + f_{\text{CUT}}
    \label{eq:new_axes}
\end{equation}
This is equivalent to placing the origin of the new integration variables $f_{1}^{'}$ at the center frequency of the CUT. As each channel becomes the CUT, such origin is moved onto its frequency. This way, we obtain:
\begin{equation}
    \begin{aligned}
        &\Delta \beta^{(n_s)}\bigl( f_1' + f_{\text{CUT}},\, f_2' + f_{\text{CUT}},\, f_{\text{CUT}} \bigr)
        = -4\pi^{2} f_1' f_2' \\
        &\quad \cdot \Biggl( \beta_2^{(n_s)} 
        + \pi \beta_3^{(n_s)} \bigl( f_1' + f_2' + 2f_{\text{CUT}} - 2f_c^{(n_s)} \bigr) \\
        &\quad + \frac{2}{3} \pi^2 \beta_4^{(n_s)} \Bigl( 
        \bigl( f_1' + f_{\text{CUT}} - f_c^{(n_s)} \bigr)^2 \\
        &\quad + \bigl( f_1' + f_{\text{CUT}} - f_c^{(n_s)} \bigr)
               \bigl( f_2' + f_{\text{CUT}} - f_c^{(n_s)} \bigr) \\
        &\quad + \bigl( f_2' + f_{\text{CUT}} - f_c^{(n_s)} \bigr)^2 
        \Bigr) \Biggr)
    \end{aligned}
\label{eq:delta_beta2_centered}
\end{equation}

This important quantity depends on the value of $\beta _{2}^{({{n}_{s}})}$ and on a `correction' that in turn depends on $\beta _{3}^{({{n}_{s}})}$, $\beta _{4}^{({{n}_{s}})}$. 
Then, as an approximation, only in the inner bracket we replace $f_{1}^{'}$ and $f_{2}^{'}$ with their center values over the integration island $x$, that is:
\begin{equation}
    f_1' \approx f_{m_{\text{ch}}}^{(n_s)} - f_{\text{CUT}}\;\;\;,\;\;\; f_2' \approx f_{k_{\text{ch}}}^{(n_s)} - f_{\text{CUT}}
\end{equation}
We get:
\begin{equation}
\begin{aligned}
  & \Delta \beta^{(n_s)} \bigl(f_1' + f_{\text{CUT}},\, f_2' + f_{\text{CUT}},\, f_{\text{CUT}}\bigr) \approx -4 \pi^2 f_1' f_2' \\
  & \cdot \Biggl(
    \beta_2^{(n_s)} 
    + \pi \beta_3^{(n_s)} \bigl[
      f_{m_{\text{ch}}}^{(n_s)} + f_{k_{\text{ch}}}^{(n_s)} - 2 f_c^{(n_s)}
    \bigr] \\
  & + \frac{2}{3} \pi^2 \beta_4^{(n_s)} \Bigl(
      \bigl(f_{m_{\text{ch}}}^{(n_s)} - f_c^{(n_s)}\bigr)^2 
      + \bigl(f_{m_{\text{ch}}}^{(n_s)} - f_c^{(n_s)}\bigr) \\
  & \cdot \bigl(f_{k_{\text{ch}}}^{(n_s)} - f_c^{(n_s)}\bigr) 
      + \bigl(f_{k_{\text{ch}}}^{(n_s)} - f_c^{(n_s)}\bigr)^2
    \Bigr)
  \Biggr)
\end{aligned}
\label{eq:delta_beta2_approx}
\end{equation}
To better understand the meaning of this approximation, we point out that the result would be exact if $\beta _{2}^{({{n}_{s}})}$ was piecewise constant over the bandwidth of each channel. So, the above approximation does account for the change over frequency of $\Delta {{\beta }^{({{n}_{s}})}}$, due to the presence of a non-zero $\beta _{3}^{({{n}_{s}})}$, $\beta _{4}^{({{n}_{s}})}$. We are simply assuming that dispersion is locally constant over each channel, a reasonable assumption given the typical bandwidth of a channel and the small values of $\beta _{3}^{({{n}_{s}})}$, $\beta _{4}^{({{n}_{s}})}$. This approximation however turns out to be crucial in achieving a closed-form solution of the integrals.
We can now define:
\begin{equation}
\begin{aligned}
    & \beta_{2,\text{eff},x}^{(n_s)} = \beta_2^{(n_s)} 
        + \pi \beta_3^{(n_s)} \bigl[
          f_{m_{\text{ch}}}^{(n_s)} + f_{k_{\text{ch}}}^{(n_s)} - 2 f_c^{(n_s)}
        \bigr] \\
      & + \frac{2}{3} \pi^2 \beta_4^{(n_s)} \Bigl(
          \bigl(f_{m_{\text{ch}}}^{(n_s)} - f_c^{(n_s)}\bigr)^2 
          + \bigl(f_{m_{\text{ch}}}^{(n_s)} - f_c^{(n_s)}\bigr) \\
      & \cdot \bigl(f_{k_{\text{ch}}}^{(n_s)} - f_c^{(n_s)}\bigr) 
          + \bigl(f_{k_{\text{ch}}}^{(n_s)} - f_c^{(n_s)}\bigr)^2
        \Bigr)
\end{aligned}
\label{eq:beta2_effective}
\end{equation}
which turns out to be a sort of `effective' value of  $\beta _{2}^{({{n}_{s}})}$, constant over the island $x$. Eq.~(\ref{eq:delta_beta2_approx}) can then be drastically simplified as:
\begin{equation}
    \begin{aligned}
        &\Delta \beta^{(n_s)}\bigl( f_1' + f_{\text{CUT}},\, f_2' + f_{\text{CUT}},\, f_{\text{CUT}} \bigr)
        \approx -4\pi^{2} f_1' f_2' \beta_{2,\text{eff},x}^{(n_s)}
    \end{aligned}
\label{eq:delta_beta2_simplified}
\end{equation}
\subsection{Loss}
We can then apply the same strategy to the loss term:
\begin{equation}
\begin{aligned}
      & \Delta {{\alpha_T }^{({{n}_{s}})}}\left( f_{1}^{'}+{{f}_{\text{CUT}}},f_{2}^{'}+{{f}_{\text{CUT}}},{{f}_{\text{CUT}}},z \right) \\
     & \approx{{\alpha_T }^{({{n}_{s}})}}\left( f_{{{m}_{\text{ch}}}}^{({{n}_{s}})},z \right)+{{\alpha_T }^{({{n}_{s}})}}\left( f_{{{k}_{\text{ch}}}}^{({{n}_{s}})},z \right)\\
     &-{{\alpha_T }^{({{n}_{s}})}}\left( {{f}_{\text{CUT}}},z \right)
     +{{\alpha_T }^{({{n}_{s}})}}\left( f_{{{m}_{\text{ch}}}}^{({{n}_{s}})}+f_{{{k}_{\text{ch}}}}^{({{n}_{s}})}-{{f}_{\text{CUT}}},z \right) \\
\end{aligned}
\label{eq:delta_alpha}
\end{equation}
Once more, the result would be exact if loss was frequency-flat over the bandwidth of each of the channels involved, a very reasonable assumption. 

\subsection{Spatial power-profile formalism }
Substituting Eqs.~(\ref{eq:new_axes}),~(\ref{eq:delta_beta2_simplified}),~(\ref{eq:delta_alpha}) into Eq.~(\ref{eq:NLI_integral_x}) yields:
\begin{equation}
    \begin{aligned} 
         & G_{\text{NLI,}x}^{(n_s)}\left( f_{\text{CUT}} \right) = \frac{16}{27} \, \Gamma^{(n_s)}\left( f_{\text{CUT}} \right) \\ 
         &\cdot G_{\text{WDM,}m_{\text{ch}}}^{(n_s)} G_{\text{WDM,}k_{\text{ch}}}^{(n_s)} 
         G_{\text{WDM,}n_{\text{ch}}}^{(n_s)} 
         \left( \gamma^{(n_s)}_x \right)^2 \\
         &\int_{f_{k_{\text{ch}}}^{(n_s)} - f_{\text{CUT}} - B_{k_{\text{ch}}}^{(n_s)}/2}^{f_{k_{\text{ch}}}^{(n_s)} - f_{\text{CUT}} + B_{k_{\text{ch}}}^{(n_s)}/2} 
         \int_{f_{m_{\text{ch}}}^{(n_s)} - f_{\text{CUT}} - B_{m_{\text{ch}}}^{(n_s)}/2}^{f_{m_{\text{ch}}}^{(n_s)} - f_{\text{CUT}} + B_{m_{\text{ch}}}^{(n_s)}/2} \\
          & \left| \rho^{(n_s)}\left(f_{1}' + f_{\text{CUT}}, f_{2}' + f_{\text{CUT}}, f_{\text{CUT}} \right) \right|^2 \, df_1' \, df_2' \\
   \end{aligned}
  \label{eq:NLI_integral_generic}
\end{equation}
with the integrand function:
\begin{equation}
\begin{aligned}
&\left| \rho^{(n_s)}\left(f_{1}' + f_{\text{CUT}}, f_{2}' + f_{\text{CUT}}, f_{\text{CUT}} \right) \right|^2 \\
&\approx e^{-2\!\int_{0}^{L_s^{(n_s)}}\!\!\alpha_T^{(n_s)}(f_{\text{CUT}}, z)\,dz} \\
& \cdot \Bigg| \int_{0}^{L_s^{(n_s)}} \!\!\! 
e^{j4\pi^2 f_1' f_2' \beta_{2,\text{eff},x}^{(n_s)} z} \cdot e^{ -\!\!\int_{0}^{z}\!\!\alpha_T^{(n_s)}(f_{m_{\text{ch}}}^{(n_s)}, z')\,dz' } \\
&\cdot e^{ -\!\!\int_{0}^{z}\!\!\alpha_T^{(n_s)}(f_{k_{\text{ch}}}^{(n_s)}, z')\,dz' } \cdot e^{ \int_{0}^{z}\!\!\alpha_T^{(n_s)}(f_{\text{CUT}}, z')\,dz' } \, dz \\
& \cdot e^{ -\!\!\int_{0}^{z}\!\!\alpha_T^{(n_s)}(f_{m_{\text{ch}}}^{(n_s)} + f_{k_{\text{ch}}}^{(n_s)} - f_{\text{CUT}}, z')\,dz' } \Bigg|^2
\end{aligned}
\label{eq:FWM_efficiency}
\end{equation}

Note that Eq.~(\ref{eq:FWM_efficiency}) is indicated as an approximation of $\mid\rho^{(n_s)}\mid^2$ because two approximated expressions, Eq.~(\ref{eq:delta_beta2_simplified}) and Eq.~(\ref{eq:delta_alpha}), are substituted into the defining equation for $\mid\rho^{(n_s)}\mid^2$ in Eq.~(\ref{eq:link_function}). The same is true for Eq.~(\ref{eq:32_pho}) below.

We now define the `normalized Spatial Power-Profile', or SPP, of the ${i}_{\text{ch}}\text{-th}$ channel along the ${n}_{s}\text{-th}$ span as:
\begin{equation}
    \begin{aligned}
     p_{i_{\text{ch}}}^{(n_s)}(z) = \frac{P_{i_{\text{ch}}}^{(n_s)}(z)}{P_{i_{\text{ch}}}^{(n_s)}(0)} 
    \end{aligned}
    \label{eq:SPP}
\end{equation}
Note that  $p_{{{i}_{\text{ch}}}}^{({{n}_{s}})}\left( 0 \right)=1$. 
We then point out that such SPP can be expressed in terms of the generalized loss/gain coefficient $\alpha_T$ as follows:
\begin{equation}
    \begin{aligned}
     p_{i_{\text{ch}}}^{(n_s)}(z)
    = e^{-2 \int_{0}^{z} \alpha_T^{(n_s)}\big( f_{i_{\text{ch}}}^{(n_s)}, z' \big)\, dz'}   
    \end{aligned}
    \label{eq:SPP_polynominal}
\end{equation}
which takes into account the accumulated loss/gain along the propagation fiber, experienced up to length $z$ into the span, due to fiber intrinsic loss, Raman amplification, ISRS and possible lumped loss along the fiber. 

Using Eq.~(\ref{eq:SPP_polynominal}) in Eq.~(\ref{eq:FWM_efficiency}), we get:
\begin{equation}
    \begin{aligned}
    &{{\left| \rho^{(n_s)} \left( f_{1}^{'}+{{f}_{\text{CUT}}},f_{2}^{'}+{{f}_{\text{CUT}}},{{f}_{\text{CUT}}} \right) \right|}^{2}}\approx p_{\text{CUT}}^{({{n}_{s}})}\left( L_{s}^{\left( {{n}_{s}} \right)} \right)\\
    &\cdot {{\left| \int\limits_{0}^{L_{s}^{\left( {{n}_{s}} \right)}}{\sqrt{\frac{p_{{{m}_{\text{ch}}}}^{({{n}_{s}})}\left( z \right)p_{{{k}_{\text{ch}}}}^{({{n}_{s}})}\left( z \right)p_{{{n}_{\text{ch}}}}^{({{n}_{s}})}\left( z \right)}{p_{\text{CUT}}^{({{n}_{s}})}\left( z \right)}}\ {{e}^{j4{{\pi }^{2}}f_{1}^{'}f_{2}^{'} \beta _{2,\text{eff},x}^{\left( {{n}_{s}} \right)}\cdot z}}\ }dz \right|}^{2}}
    \label{eq:32_pho}
    \end{aligned}
\end{equation}

This form of the integrand function departs substantially from what has been so far used or proposed for all other CFMs in the literature. Typically, from Eq.~(\ref{eq:FWM_efficiency}) onward, a completely different approach at performing calculations was used, focusing on the loss/gain coefficient $\alpha_T$. Here instead, we focus on the SPP of each channel. 

We then express the combination of the SPPs under square root in Eq.~(\ref{eq:32_pho}) as a single function which depends on the island $x$  being addressed, $p_x^{(n_s)}(z)$, as follows:
\begin{equation}
    \begin{aligned}           
    p_x^{(n_s)}(z) = \sqrt{\frac{
    p_{m_{\text{ch}}}^{(n_s)}(z)\,
    p_{k_{\text{ch}}}^{(n_s)}(z)\,
    p_{n_{\text{ch}}}^{(n_s)}(z)
    }{
    p_{\text{CUT}}^{(n_s)}(z)
    }}
    \label{eq:rho_general}
    \end{aligned}
\end{equation}

Using Eqs.~(\ref{eq:32_pho}), (\ref{eq:rho_general}) in Eq.~(\ref{eq:NLI_integral_generic}) we obtain the expression for the NLI PSD contribution of any generic island $x$ (approximated as a rectangle, see Fig.~\ref{fig:rectangles}) in any generic span $n_s$:
\begin{equation}
    \begin{aligned} 
         & G_{\text{NLI,}x}^{(n_s)}\left( f_{\text{CUT}} \right) = \frac{16}{27} \, \Gamma^{(n_s)}\left( f_{\text{CUT}} \right) \\ 
         &\cdot G_{\text{WDM,}m_{\text{ch}}}^{(n_s)} G_{\text{WDM,}k_{\text{ch}}}^{(n_s)} 
         G_{\text{WDM,}n_{\text{ch}}}^{(n_s)} 
         \left( \gamma^{(n_s)}_x \right)^2 p_{\text{CUT}}^{({{n}_{s}})}\left( L_{s}^{\left( {{n}_{s}} \right)} \right)\\
         &\int_{f_{k_{\text{ch}}}^{(n_s)} - f_{\text{CUT}} - B_{k_{\text{ch}}}^{(n_s)}/2}^{f_{k_{\text{ch}}}^{(n_s)} - f_{\text{CUT}} + B_{k_{\text{ch}}}^{(n_s)}/2} 
         \int_{f_{m_{\text{ch}}}^{(n_s)} - f_{\text{CUT}} - B_{m_{\text{ch}}}^{(n_s)}/2}^{f_{m_{\text{ch}}}^{(n_s)} - f_{\text{CUT}} + B_{m_{\text{ch}}}^{(n_s)}/2} \\
          & \left| \int_0^{L_s^{(n_s)}}
    p_x^{(n_s)}(z) \,
    e^{j 4 \pi^2 f_1' f_2' \beta_{2,\text{eff},x}^{(n_s)} z}
    \, dz \right|^2 \, df_1' \, df_2' \\
   \end{aligned}
  \label{eq:NLI_integral_generic_pch}
\end{equation}

This important formula shows that the key analytical hurdle to achieving a CFM is solving for the `core' integrals of Eq.~(\ref{eq:NLI_integral_generic_pch}), which we call $K_x^{(n_s)} (f_{\text{CUT}})$:
\begin{equation}
\begin{aligned}
    K_x^{(n_s)}& (f_{\text{CUT}}) = 
    \int_{f_{k_{\text{ch}}}^{(n_s)} - f_{\text{CUT}} - {B_{k_{\text{ch}}}^{(n_s)}}/{2}}^{
                  f_{k_{\text{ch}}}^{(n_s)} - f_{\text{CUT}} + {B_{k_{\text{ch}}}^{(n_s)}}/{2}}
    \int_{f_{m_{\text{ch}}}^{(n_s)} - f_{\text{CUT}} - {B_{m_{\text{ch}}}^{(n_s)}}/{2}}^{
                  f_{m_{\text{ch}}}^{(n_s)} - f_{\text{CUT}} + {B_{m_{\text{ch}}}^{(n_s)}}/{2}}  \\ 
    & \left| \int_0^{L_s^{(n_s)}}
    p_x^{(n_s)}(z) \,
    e^{j 4 \pi^2 f_1' f_2' \beta_{2,\text{eff},x}^{(n_s)} z}
    \, dz \right|^2df_1' df_2'
\label{eq:integral_core}
\end{aligned}
\end{equation}

Henceforth we will focus on analytically dealing with these core integrals.

\subsection{Integrals for SCI and XCI}
The SCI and XCI islands are the red and yellow islands, respectively, in Fig.~\ref{fig:integration_domain}. As  discussed in Sect.~(\ref{subsect:NLI_PSD}), we have approximated each lozenge-shaped island with the rectangle or square that inscribes it, as shown in Fig.~\ref{fig:rectangles}. This approximation is already reflected in the integration domain of the `core' frequency integrals Eqs.~(\ref{eq:NLI_integral_generic_pch}) and (\ref{eq:integral_core}).

Note that, if channels had non-uniform spacing and non-identical bandwidths, the corresponding island layout would be much less regular than shown in Fig.~\ref{fig:integration_domain}. See \cite{2025_TechRxiv_Poggiolini} for a few examples of islands in these cases. However, it is always possible to inscribe an island into a square and therefore the following results are completely general. They can deal with non-uniform combs too, with no change.

\subsubsection{XCI contribution}
We first focus on the yellow XCI islands straddling the horizontal $f'_1$ axis, i.e., the axis with ${{f}'_{2}}=0$.
They are found by imposing $k_{\text{ch}}^{(n_s)}=n_{\text{CUT}}^{(n_s)}$ and  $m_{\text{ch}}^{(n_s)}=n_{\text{ch}}^{(n_s)}$, resulting in island identifiers of the form:
\begin{equation}
x=\left({n_{\text{ch}}^{(n_{s})}, n_{\text{CUT}}^{(n_{s})}, n_{\text{ch}}^{(n_{s})}}\right)
\end{equation}
with ${n}_{\text{ch}}=1 \ldots N_{\rm ch}$, and ${n}_{\text{ch}}\ne {n}_{\text{CUT}}$ because ${n}_{\text{ch}}= {n}_{\text{CUT}}$ would be SCI.
We call $ \mathcal{X}_{\rm{\rm XCI}_h}$ the set of these `horizontal axis' XCI islands. 

Focusing on these islands in Eqs.~(\ref{eq:NLI_integral_generic_pch}) and (\ref{eq:integral_core}), we obtain: 
\begin{equation}
\begin{aligned}
     & G_{x\in\mathcal{X}_{\rm{\rm XCI}_h}}(f_{\text{CUT}}) =\frac{16}{27} \cdot p_{\text{CUT}}^{(n_s)}(L_s^{(n_s)})\cdot\Gamma^{(n_s)}(f_{\text{CUT}}) \\
    & \qquad \cdot G_{\text{CUT}}^{(n_s)} \cdot\bigl(G_{\text{WDM,}n_{\text{ch}}}^{(n_s)}\bigr)^2 
    \cdot \left(  \gamma_{x\in\mathcal{X}_{\rm{\rm XCI}}}^{(n_s)}\right)^2
     \cdot K_{x\in\mathcal{X}_{\rm{\rm XCI}_h}}^{(n_s)} \\
\end{aligned}
\label{eq:GXCI}
\end{equation}
\begin{equation}
\begin{aligned}
    & K_{x\in\mathcal{X}_{\rm{\rm XCI}_h}}^{(n_s)}  =     
    \int_{-B_{\text{CUT}}/2}^{B_{\text{CUT}}/2}
    \int_{f_{n_{\text{ch}}}^{(n_s)} - f_{\text{CUT}} - {B_{n_{\text{ch}}}^{(n_s)}}/{2}}^{
                  f_{n_{\text{ch}}}^{(n_s)} - f_{\text{CUT}} + {B_{n_{\text{ch}}}^{(n_s)}}/{2}}  \\ 
    & \left| \int_0^{L_s^{(n_s)}}
    p_{x\in\mathcal{X}_{\rm{\rm XCI}_h}}^{(n_s)} (z) \,
    e^{j 4 \pi^2 f_1' f_2' \beta_{2,\text{eff},{n_{\text{ch}}}}^{(n_s)} z}
    \, dz \right|^2  df_1' df_2'
\label{eq:XCI_h_integral_core}
\end{aligned}
\end{equation}
where the nonlinear coefficient, specializing the general expression Eq.~(\ref{eq:gamma_approximation}) to the XCI case, can be written:
\begin{align}
  \gamma^{(n_s)}_{x\in\mathcal{X}_{\rm{\rm XCI}}}= \frac{2\pi f_{\text{CUT}}}{c} \cdot \frac{2n_2}{A_{\text{eff}} (f_{\text{CUT}}) + A_{\text{eff}} (f_{n_{\text{ch}}}^{(n_s)})} 
  \label{eq:gamma_XCI}
\end{align}
The quantity $A_{\rm eff}(f)$ is the effective area of the fiber at the frequency $f$, for which we used \cite{2022_OE_Shevchenko} Eq.~(4) (derived from \cite{1977_BellTech_Marcuse}). 
Importantly, the power-profile factor $p_{x}^{(n_{s})}(z)$ becomes simply the SPP of the channel with index $n_{\rm ch}$, once the islands indices $x\in \mathcal{X}_{\rm{\rm XCI}_h}$ are plugged into Eq.~(\ref{eq:rho_general}) :
\begin{equation}
    \begin{aligned}           
      p_{x\in\mathcal{X}_{\rm{\rm XCI}_h}}^{(n_s)}(z) = \sqrt{\frac{
    p_{n_{\text{ch}}}^{(n_s)}(z)\,
    p_{n_{\text{CUT}}}^{(n_s)}(z)\,
    p_{n_{\text{ch}}}^{(n_s)}(z)
    }{
    p_{n_{\text{CUT}}}^{(n_s)}(z)
    }}=  p_{n_{\text{ch}}}^{(n_s)}(z)
    \label{eq:34_G_NLI with Kx}
    \end{aligned}
\end{equation}
This way, the core integrals simply become:
\begin{equation}
\begin{aligned}
    &  K_{x\in\mathcal{X}_{\rm{\rm XCI}_h}}^{(n_s)} =     
    \int_{-B_{\text{CUT}}/2}^{B_{\text{CUT}}/2}
    \int_{f_{n_{\text{ch}}}^{(n_s)} - f_{\text{CUT}} - {B_{n_{\text{ch}}}^{(n_s)}}/{2}}^{
                  f_{n_{\text{ch}}}^{(n_s)} - f_{\text{CUT}} + {B_{n_{\text{ch}}}^{(n_s)}}/{2}}  \\ 
    & \left| \int_0^{L_s^{(n_s)}}
    p_{n_{\text{ch}}}^{(n_s)}(z) \,
    e^{j 4 \pi^2 f_1' f_2' \beta_{2,\text{eff},{n_{\text{ch}}}}^{(n_s)} z}
    \, dz \right|^2  df_1' df_2'
\label{eq:XCI_h_integral_core_SPP}
\end{aligned}
\end{equation}

We then focus on the yellow XCI islands straddling the vertical $f'_2$ axis, i.e., the axis with ${{f}'_{1}}=0$.
They are found by imposing $m_{\text{ch}}^{(n_s)}=n_{\text{CUT}}^{(n_s)}$ and  $k_{\text{ch}}^{(n_s)}=n_{\text{ch}}^{(n_s)}$, resulting in island identifiers of the form:
\begin{equation}
x=\left({n_{\text{CUT}}^{(n_{s})}, n_{\text{ch}}^{(n_{s})}, n_{\text{ch}}^{(n_{s})}}\right)
\end{equation}
again with ${n}_{\text{ch}}=1 \ldots N_{\rm ch}$ and ${n}_{\text{ch}}\ne {n}_{\text{CUT}}$.
We call $ \mathcal{X}_{\rm{\rm XCI}_v}$ the set of these `vertical axis' XCI islands. 

Going through the same procedure used in the case of the horizontal axis XCI islands, we get:
\begin{equation}
\begin{aligned}
   & K_{x\in\mathcal{X}_{\rm{\rm XCI}_v}}^{(n_s)}  =     
     \int_{-B_{\text{CUT}}/2}^{B_{\text{CUT}}/2}
     \int_{f_{n_{\text{ch}}}^{(n_s)} - f_{\text{CUT}} - {B_{n_{\text{ch}}}^{(n_s)}}/{2}}^{
                  f_{n_{\text{ch}}}^{(n_s)} - f_{\text{CUT}} + {B_{n_{\text{ch}}}^{(n_s)}}/{2}}              
    \\ 
    & \left| \int_0^{L_s^{(n_s)}}
    p_{n_{\text{ch}}}^{(n_s)}(z) \,
    e^{j 4 \pi^2 f_1' f_2' \beta_{2,\text{eff},{n_{\text{ch}}}}^{(n_s)} z}
    \, dz \right|^2  df_2' df_1'
\label{eq:XCI_v_integral_core_SPP}
\end{aligned}
\end{equation}

\begin{figure}
    \centering
    \includegraphics[width=0.7\linewidth]{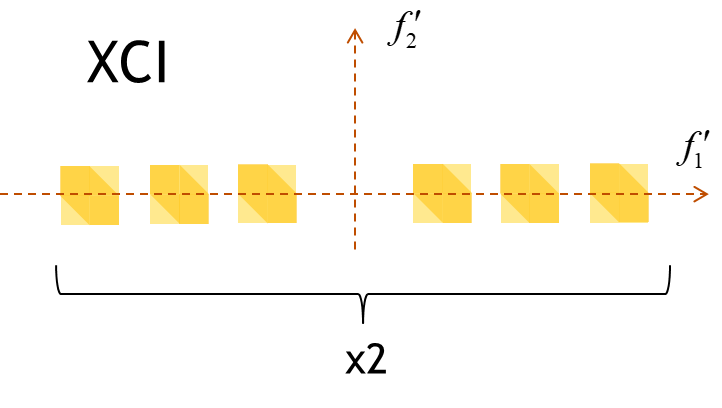}
    \caption{With reference to the example of Fig.~\ref{fig:rectangles}, these are the XCI islands on the horizontal axis that need to be evaluated to obtain the overall XCI contribution, provided that the result is multiplied by two to account for the equal contribution from the vertical XCI islands.}
    \label{fig:XCI}
\end{figure}

The only difference between Eq.~(\ref{eq:XCI_v_integral_core_SPP}) and Eq.~(\ref{eq:XCI_h_integral_core_SPP}) is the inverted order of the differentials $df'_1$ and $df'_2$. However, a simple change of variables, i.e., swapping $f'_1$ with $f'_2$, shows that, given $n_{\text{CUT}}^{(n_{s})}$ and $n_{\text{ch}}^{(n_{s})}$, the result of Eq.~(\ref{eq:XCI_v_integral_core_SPP}) and Eq.~(\ref{eq:XCI_h_integral_core_SPP}) are identical. This implies that, to account for all of XCI, we only need to carry out the integrals on, for instance, the horizontal axis XCI islands and then multiply the result by two, as depicted in Fig.~\ref{fig:XCI} with reference to the example of Fig.~\ref{fig:rectangles}. 

This also means that we can merge together Eqs.~(\ref{eq:XCI_h_integral_core_SPP}) and (\ref{eq:XCI_v_integral_core_SPP}) and simplify the notation for the XCI equations as follows:
\begin{equation}
\begin{aligned}
     &\quad G_{\text{XCI,}n_{\text{ch}}}^{(n_s)} (f_{\text{CUT}}) =\frac{32}{27} \cdot p_{\text{CUT}}^{(n_s)}(L_s^{(n_s)})\cdot\Gamma^{(n_s)}(f_{\text{CUT}}) \\
    & \qquad \qquad \cdot G_{\text{CUT}}^{(n_s)} \cdot\bigl(G_{\text{WDM,}n_{\text{ch}}}^{(n_s)}\bigr)^2 
    \cdot \left(  \gamma_{\text{XCI,}n_{\text{ch}}}^{(n_s)} \right)^2
     \cdot K_{\text{XCI,}n_{\text{ch}}}^{(n_s)}  \\
\end{aligned}
\label{eq:GXCI_2}
\end{equation}
\begin{equation}
\begin{aligned}
    K_{\text{XCI,}n_{\text{ch}}}^{(n_s)}  & =     
    \int_{-B_{\text{CUT}}/2}^{B_{\text{CUT}}/2}
    \int_{f_{n_{\text{ch}}}^{(n_s)} - f_{\text{CUT}} - {B_{n_{\text{ch}}}^{(n_s)}}/{2}}^{
                  f_{n_{\text{ch}}}^{(n_s)} - f_{\text{CUT}} + {B_{n_{\text{ch}}}^{(n_s)}}/{2}}  \\ 
    & \left| \int_0^{L_s^{(n_s)}}
    p_{n_{\text{ch}}}^{(n_s)}  (z) \,
    e^{j 4 \pi^2 f_1' f_2' \beta_{2,\text{eff},{n_{\text{ch}}}}^{(n_s)} z}
    \, dz \right|^2  df_1' df_2'
\label{eq:XCI_integral_core}
\end{aligned}
\end{equation}
where we recognize that the only index needed to identify the XCI islands is $n_{\text{ch}}^{(n_{s})}$. Each contribution physically represents the XCI produced by the WDM channel $n_{\text{ch}}^{(n_{s})}$ onto the CUT. Also, $\gamma_{\text{XCI,}n_{\text{ch}}}^{(n_s)}$ coincides with Eq.~(\ref{eq:gamma_XCI}).

\begin{figure}
    \centering
    \includegraphics[width=0.75\linewidth]{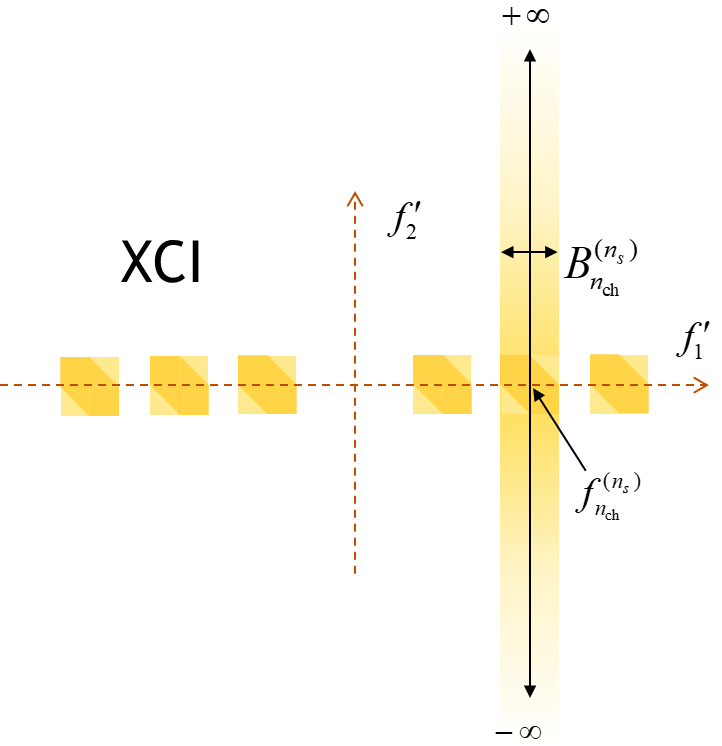} \vspace{5mm}
    \includegraphics[width=0.9\linewidth]{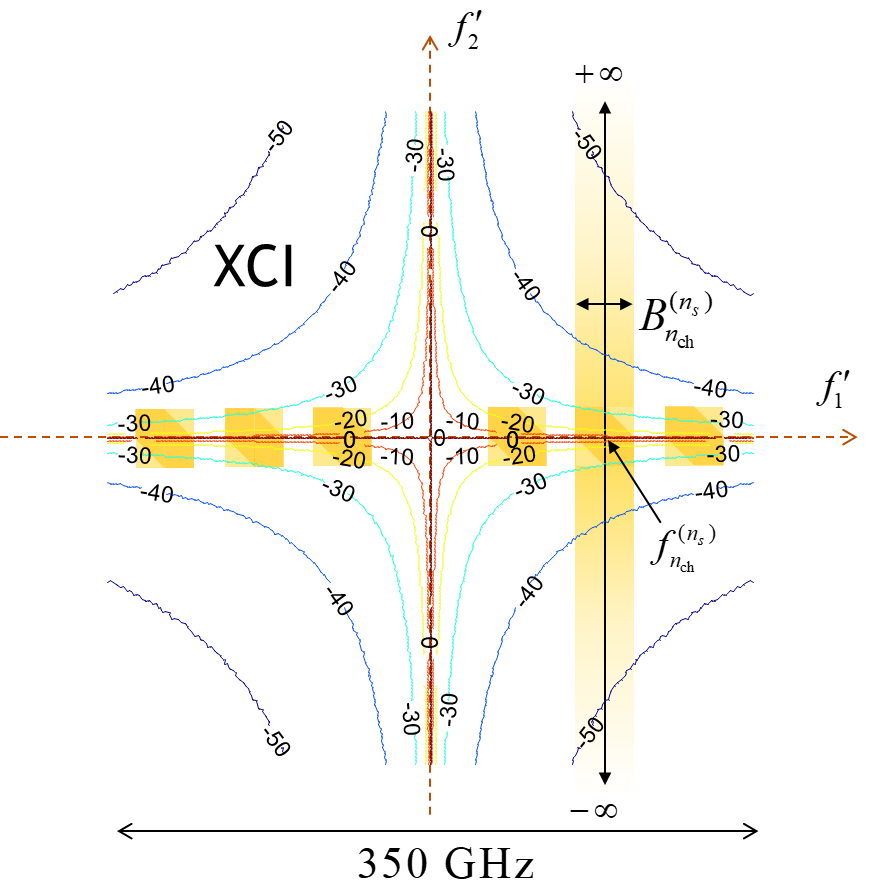}
    \caption{\textbf{Top:} the XCI island centered at $f_{{{n}_{\text{ch}}}}^{({{n}_{s}})}$ along the ${{f}_{1}}$ axis is stretched vertically to $\left[ -\infty ,\infty  \right]$. \textbf{Bottom:} same as above, with superimposed the contour lines (in dB) of an example of the integrand function which results from assuming a single span of SMF of length 100km. The integrand function achieves its maximum on the axes and is normalized to 0 dB there. The islands are drawn assuming 7 channels at 28 GBaud with 50 GHz spacing. It is apparent that the stretched island adds a region of the plane where the integrand function is far attenuated. The same approximation is done for all islands, with similar justification.}
    \label{fig:big_mess}
\end{figure}

\subsection{SPP polynomial representation}
The next step towards reaching a CFM is to express the SPP $p_{n_{\text{ch}}}^{(n_s)}(z)$ of the $n_{\text{ch}}^{(n_s)}$-th channel as a polynomial in the distance variable $z$:
\begin{equation}       
    p_{n_{\text{ch}}}^{(n_s)}(z) =  \sum_{n=0}^{N_p} p_{n,{n_{\text{ch}}}}^{(n_s)} z^n 
    \label{eq:the_polynomial}
\end{equation}
This step, to the best of our knowledge, has not been previously tried in the context of CFMs. The key aspect of this representation is that it allows fully closed-form solutions for the core NLI integrals to be found, for \textit{all types} of islands. 

These closed-form solutions may however become rather complex, specifically as the polynomial degree $N_{p}$ is increased. In the next section, we present instead a remarkably simple general closed-form solution for XCI, of limited complexity for any polynomial degree $N_{p}$, which, however, requires one further approximation. Such approximation causes negligible error in systems with highly dispersive fibers and high symbol rates (see below) and therefore it covers the majority of modern systems. We will also discuss in detail accuracy vs.~$N_p$ in Sects.~\ref{sec:SPP_polyfit} and \ref{sec:err_vs_Np}.

\subsubsection{The XCI contribution}
To achieve a  simple general closed-form solution for XCI, a further approximation is needed.
The approximation consists of stretching the XCI islands to $\left[ -\infty ,\infty  \right]$ in the dimension where their size would be $\left[ -{{{B}_{\text{CUT}}}}/{2}\;,{{{B}_{\text{CUT}}}}/{2}\; \right]$. An example is shown in Fig.~\ref{fig:big_mess}. 
This may seem a quite substantial approximation but it is very common in CFMs since it typically leads to negligible error in modern high-capacity, long-haul SMF systems. 

In Fig.~\ref{fig:big_mess}-bottom, the contour lines of the integrand (resulting from the inner integral in $z$ in Eq.~(\ref{eq:XCI_h_integral_core_SPP})) are shown, in dB, normalized to 0~dB at their maximum, which occurs all along the axes, assuming 100 km of SMF. Note that it is essentially dispersion, and specifically $\beta_{2,\text{eff},{n_{\text{ch}}}}^{(n_s)}$, that sets the decay of the integrand moving away from the axes. The more dispersion, the faster the decay. In the figure, the channels are relatively low-rate: 28~GBaud, with spacing 50 GHz. The plot clearly shows that at the edge of the stretched island, a decay of $-30$~dB has already occurred. If we consider the next island closer to the origin, even though less decay has occurred at its edges, it is still a decay greater than $-20$~dB. So, stretching upward that island does not cause major error either. The argument gets stronger as one moves away from the origin.

Note that this general feature of the integrand was originally pointed out already in \cite{2012_JLT_Poggiolini}, and has since been exploited in many papers to achieve CFMs of various types. Whether this approximation is acceptable or not, it  depends on the system parameters. With low dispersion fibers, the decay of the integrand is slower. But also the vertical size of the island before stretching is important, which is set by the CUT symbol rate. As a reasonable guideline, it must be that the product  $\beta_{2,\text{eff},{n_{\text{ch}}}}^{(n_s)}\cdot R_{\rm CUT}^2>0.01$ 1/km. This is achieved for instance for $\beta_{2,\text{eff},{n_{\text{ch}}}}^{(n_s)}=20$ ps$^2$/km (typical for G.652 SMF) and $R_{\rm CUT}=0.028$~TBaud (or 28 GBaud). With modern systems trending towards 100+~GBaud, even fibers with dispersion as low as 2~ps${^2}$/km could be dealt with. However, if dispersion is even lower, or otherwise symbol rates are very low such as in multi-subcarrier systems, then these scenarios fall outside of the scope of this approximation. In that case, the island-stretching approximation must be avoided and PCFM allows that, resulting in more complex but still fully closed-form formulas. A presentation of these results is however left for a separate submission.

Quite remarkably, once the stretching of the islands has been performed, the XCI kernel integrals admit a compact fully closed-form exact solution, with no further approximation:
\begin{equation}
\begin{aligned}
K_{\text{XCI,}n_{\text{ch}}}^{(n_s)} 
&= \frac{L_s^{(n_s)}}{2\pi |\beta_{2,\text{eff},{n_{\text{ch}}}}^{(n_s)}|} \cdot
\left| \log \left( \frac{f_{n_{\text{ch}}}^{(n_s)} - f_{\text{CUT}} + {B_{n_{\text{ch}}}^{(n_s)}}/{2}}
{f_{n_{\text{ch}}}^{(n_s)} - f_{\text{CUT}} - {B_{n_{\text{ch}}}^{(n_s)}}/{2}} \right) \right|  \\
&\cdot \sum_{n=0}^{N_p} \sum_{k=0}^{N_p} \frac{p_{n,n_{\text{ch}}}^{(n_s)} p_{k,n_{\text{ch}}}^{(n_s)} \left( L_s^{(n_s)} \right)^{(n+k)}}{n+k+1} 
\end{aligned}
\label{eq:CF_KXCI}
\end{equation}

Notice that the summations are finite and run over the degree of the polynomial that is needed to describe the SPP. We investigated what degrees are typically needed. The results are shown in Sects.~\ref{sec:SPP_polyfit} and \ref{sec:err_vs_Np}.

\subsubsection{The SCI contribution}
\label{sec:SCI_core}

The SCI integration takes place over the red islands represented in the examples of Fig.~\ref{fig:integration_domain}. Formally, it corresponds to the triple: 
\begin{equation}
x=\left(n_{\text{CUT}}^{(n_{s})}, n_{\text{CUT}}^{(n_{s})}, n_{\text{CUT}}^{(n_{s})}\right)
\end{equation}

As in the case of XCI, we first approximate the SCI lozenge with the square that inscribes it, as shown in Fig.~\ref{fig:rectangles}. Unlike the XCI case, here it is not possible to stretch the integration domain to infinity, along either one of the axes. The reason is that integration to infinity straddling either one of the axes inevitably results in divergence. The SCI core integrals therefore need to be integrated without any further approximations.

In the SCI island, we have:
\begin{align*}
  & f_{m_{\text{ch}}}^{(n_s)} = f_{k_{\text{ch}}}^{(n_s)} = f_{n_{\text{ch}}}^{(n_s)} = f_{\text{CUT}}\\
  & B_{m_{\text{ch}}}^{(n_s)} = {B_{k_{\text{ch}}}^{(n_s)}} = {B_{n_{\text{ch}}}^{(n_s)}} = B_{\text{CUT}}
\end{align*}
Therefore, taking the above into account in Eqs.~(\ref{eq:NLI_integral_generic_pch}) and (\ref{eq:integral_core}), we get for SCI:
\begin{equation}
\begin{aligned}
 \quad G_{\text{SCI}}^{(n_s)}(f_{\text{CUT}}) &=\frac{16}{27} \cdot p_{\text{CUT}}^{(n_s)}(L_s^{(n_s)}) \cdot \Gamma^{(n_s)}(f_{\text{CUT}}) \\ &\ \cdot \bigl(G_{\text{CUT}}^{(n_s)}\bigr)^3 \cdot \left| \gamma^{(n_s)}_{\text{SCI}}\right|^2
 \cdot K_{\text{SCI}}^{(n_s)} \\
\label{eq:GXCI_3}
\end{aligned}
\end{equation}
\begin{equation}
\begin{aligned}
    K_{\text{SCI}}^{(n_s)} & = 
    \int_{-B_{\text{CUT}}/2}^{B_{\text{CUT}}/2}
    \int_{-B_{\text{CUT}}/2}^{B_{\text{CUT}}/2}  df_1' df_2'\\ 
    & \left| \int_0^{L_s^{(n_s)}}
    p_{\text{SCI}}^{(n_s)}(z) \,
    e^{j 4 \pi^2 f_1' f_2' \beta_{2,\text{eff},{\text{CUT}}}^{(n_s)} z}
    \, dz \right|^2
\label{eq:SCI_integral_core}
\end{aligned}
\end{equation}

Similarly to the case of XCI, the SPP-related factor $p_{x}^{n_{s}}(z)$ of Eq.~(\ref{eq:rho_general}) gets drastically simplified to just the SPP of the CUT, which we then express here too as a polynomial:
\begin{equation}
    \begin{aligned}           
    p_{\text{SCI}}^{(n_s)}(z) = p_{{\text{CUT}}}^{(n_s)}(z) = \sum_{n=0}^{N_p} p_{n,{\text{CUT}}}^{(n_s)} z^n 
    \end{aligned}
\end{equation}

Quite remarkably, the core integral $K_{\text{SCI}}^{(n_s)}$ can be fully integrated analytically and exactly. However, without the possibility to resort to the stretching of the island towards infinity, the expressions for SCI turn out to be rather complex. So far we have been able to derive the full expressions for ${{N}_{p}}$ ranging from 0 to 9. A general parametric form vs.~$N_{p}$ is not yet available. You think it is likely possible and we are actively researching it.

In Eq.~(\ref{eq:K_SCI_Np0_Np1}) we show as examples the individual formulas for ${{N}_{p}}=0,1,2,3$. Note that these formulas could be optimized for computational efficiency, such as by appropriate factorization. Different strategies are available to do so, but we will not investigate them here. Also notice the presence of the hypergeometric function $_{2}F_{3}$ in the formulas. Computationally, it is not a problem since \textit{a single evaluation of it} is needed for each island, independently of the value of $N_{p}$. Also, it is implemented in Matlab, which  typically takes less than 1~ms to compute it on a conventional PC.

Higher-${{N}_{p}}$ formulas, such as for ${{N}_{p}}=5,9$, which would be too long to reproduce here, are reported in \cite{2025_TechRxiv_Poggiolini}. For a detailed discussion of the appropriate value of  ${{N}_{p}}$ please see Sects.~\ref{sec:SPP_polyfit} and \ref{sec:err_vs_Np}.

\begin{figure*}  
\begin{equation}
    \begin{aligned}
    & B = {B_{{\rm{CUT}}}}, \quad L = L_{\rm{s}}^{(n_s)},   \quad  p_n = p_{n,\text{CUT}}^{({{n}_{s}})}, \quad b_2 = \beta_{2,\text{eff,CUT}}^{\left( {{n}_{s}} \right)}, \quad x = {\pi^2}b_2{B^2}L, \quad {\rm{S}} = \sin \left( x \right), \\[3mm]
    & {\rm{C}}  = \cos \left( x \right),  \quad {\rm{SI}} = {\rm{SinIntegral}}\left( x \right), \quad {\rm{H}} = {}_2{F_3}\left( \left\{ \frac{1}{2}, \frac{1}{2} \right\}, \left\{ \frac{3}{2}, \frac{3}{2}, \frac{3}{2} \right\}, - \frac{1}{4}{{\left[ x \right]}^2} \right)  \\[3mm]
    & \left. K_{\text{SCI}}^{(n_{s})} \right|_{N_{p}=0} 
    = 2 B^{2} L^{2} p_{0}^{2} {\rm{H}} 
    + \frac{2 p_{0}^{2} (1 - {\rm{C})}}{\pi^{4} b_{2}^{2} B^{2}} 
    - \frac{2 L p_{0}^{2} {\rm{SI}}}{\pi^{2} b_{2}} \\
        \\[1mm]
    &{{\left. K_{\text{SCI}}^{({{n}_{s}})} \right|}_{{{N}_{p}}=1}} = 1 / ( 9 \pi^8 b_2^4 B^6 )\bigg[ 2 p_1^2 + 9 \pi^4 b_2^2 B^4 (2p_0^2 + 2 L p_0 p_1 + L^2 p_1^2) \\
    & \qquad -2\big(p_1^2 + \pi^4 b_2^2 B^4 (9 p_0^2 + 9 L p_0 p_1 + 4 L^2 p_1^2) \big) {\rm{C}} + 6 \pi^8 b_{2}^4 B^8 L^2 (3 p_0^2 + 3 L p_0 p_1 + L^2 p_1^2) {\rm{H}} \\
    & \qquad -2 \pi^2 b_2 B^2 L p_1^2 {\rm{S}} -2 \pi^6 b_2^3 B^6 L(9 p_0^2 + 9 L p_0 p_1 + 4 L^2 p_1^2) {\rm{SI}} \bigg]\\
    \\[1mm]
    &{{\left. K_{\text{SCI}}^{({{n}_{s}})} \right|}_{{{N}_{p}}=2}} = 1/(450 \pi^{12} b_2^6 B^{10}) \bigg[144p_2^2 + 100 \pi^4 b_2^2 B^4 (p_1^2 - 4 p_0 p_2) + 450 \pi^8 b_2^4 B^8 (2p_0^2 + 2L p_0 p_1 \\
    & \qquad   + L^2 p_1^2  + 2 L^2 p_0 p_2 + 2 L^3 p_1 p_2 + L^4  p_2^2) + \Big( -144 p_2^2 - 4 \pi^4 b_2^2 B^4 (25 p_1^2 - 100 p_0 p_2 - 18 L^2 p_2^2) \\
    & \qquad   - \pi^8 b_2^4 B^8 (900 p_0^2 + 900 L p_0 p_1 + 400 L^2 p_1^2
        + 650 L^2 p_0 p_2 + 675 L^3 p_1 p_2 + 306 L^4 p_2^2) \Big) {\rm{C}} \\
    & \qquad  + 30 \pi^{12} b_2^6 B^{12} L^2 \big(30 p_0^2+ 30 L p_0 p_1
    + 10 L^2 (p_1^2 + 2 p_0 p_2) + 15 L^3 p_1 p_2+ 6 L^4 p_2^2\Big) {\rm{H}} \\
    & \qquad  + \pi^2 b_2 B^2 L \Big(-144 p_2^2 - \pi^4 b_2^2 B^4 (100 p_1^2 + 50 p_0 p_2 + 225 L p_1 p_2 + 126 L^2 p_2^2) \Big) {\rm{S}} \\
    & \qquad  + \pi^{10} b_2^{5} B^{10}  L (-900 p_0^2 - 900 L p_0 p_1 - 400 L^2 p_1^2 - 650 L^2 p_0 p_2 - 675 L^3 p_1 p_2 - 306 L^4 p_2^2 \Big) {\rm{SI}} \bigg]\\
    \\[1mm]
    &{{\left. K_{\text{SCI}}^{({{n}_{s}})} \right|}_{{{N}_{p}}=3}} = 1/(22050 \pi^{16} b_2^8 B^{14})\Bigg[32400 p_3^2 + 7056 \pi^4 b_2^2 B^4 ( p_2^2 - 3 p_1 p_3) + 2450 \pi^8 b_2^4 B^8 (2 p_1^2 \\
    & \qquad  - 8 p_0 p_2 - 12 L p_0 p_3  - 6 L^2 p_1 p_3 - 4 L^3 p_2 p_3 - 3 L^4 p_3^2 ) + 22032 \pi^{12} b_2^6 B^{12} (2 p_0^2 + 2 L p_0 p_1  \\
    & \qquad  + L^2 p_1^2 +2 L^2 p_0 p_2 + 2 L^3 p_1 p_2 + L^4 p_2^2 + 2 L^3 p_0 p_3 + 2 L^4 p_1 p_3  + 2 L^5 p_2 p_3 + L^6 p_3^2 )\\
    & \qquad  + \Big(-32400 p_3^2 + 24 \pi^4 b_2^2 B^4 ( -294 p_2^2 + 883 p_1 p_3 + 675  L^2 p_3^2) + 4  \pi^8 b_2^4 B^8 ( -1225 p_1^2 \\
    & \qquad  + 4900 p_0 p_2 + 884 L^2 p_2^2 + 7335 L p_0 p_3 + 1029 L^2 p_1 p_3 + 2450 L^3 p_2 p_3 + 1500 L^4 p_3^2 )  \\ 
    & \qquad  + \pi^{12}  b_2^6 B^{12} ( -44100 p_0^2 - 44100 L p_0 p_1 - 19600 L^2 p_1^2 - 31875 L^2 p_0 p_2 - 33000 L^3 p_1 p_2  \\  
    & \qquad  - 15024 L^4 p_2^2 - 25725 L^3 p_0 p_3 - 28524 L^4 p_1 p_3 - 26952 L^5 p_2 p_3 - 12444 \, L^6 p_3^2 )\Big) {\rm{C}} \\
    & \qquad  + 210 \pi^{16} b_2^8 B^{16} L^2 (210 p_0^2 + 210 L p_0 p_1 + 70 L^2 p_1^2 + 140 L^2 p_0 p_2 + 105 L^3 p_1 p_2 + 42 L^4 p_2^2  \\
    & \qquad  + 105 L^3 p_0 p_3 + 84 L^4 p_1 p_3 + 70 L^5 p_2 p_3 + 30 L^6 p_3^2)  \rm{H} + \Big(- 32400 \pi^2 b_2 B^2 L p_3^2 \\
    & \qquad  +  24 \pi^6 b_2^3 B^6  L ( -294 p_2^2 + 883 p_1 p_3 + 225 L^2 p_3^2) + \pi^{10} b_2^5 B^{10} L ( -4900 p_1^2 - 2450 p_0 p_2 \\
    & \qquad  - 11025 L p_1 p_2 - 6174 L^2 p_2^2 - 3675 L p_0 p_3 - 10884 L^2 p_1 p_3 - 12252 L^3 p_2 p_3 - 6156 L^4 p_3^2 \Big) {\rm{S}} \\   
    & \qquad  + 3 \pi^{14} b_2^7  B^{14} L\Big(- 14696 p_0^2- 14696 L p_0 p_1 - L^2 (6536 p_1^2 + 10620 p_0 p_2) - L^3 (11020 p_1 p_2  \\
    & \qquad  + 8575 p_0 p_3) - L^4 (5004 p_2^2 + 9508 p_1 p_3) - 8984 L^5 p_2 p_3 - 4148 L^6 p_3^2\Big) {\rm{SI}}\Bigg]
    \end{aligned}
    \label{eq:K_SCI_Np0_Np1}
\end{equation}
\end{figure*}

\begin{figure*}
    \centering
    \includegraphics[width=0.7\linewidth]{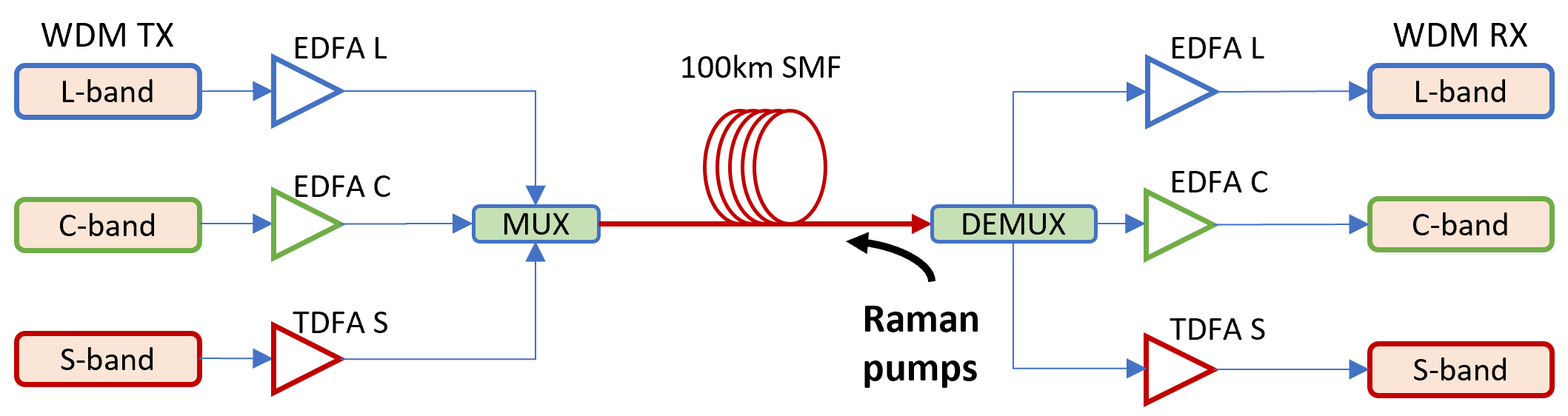}
    \caption{Schematic of the 100~km C+L+S system with backward Raman pumps.}
    \label{fig:schematic}
    \hspace{0.2cm}
\end{figure*}
\begin{figure}
    \centering
    \includegraphics[width=0.9\linewidth]{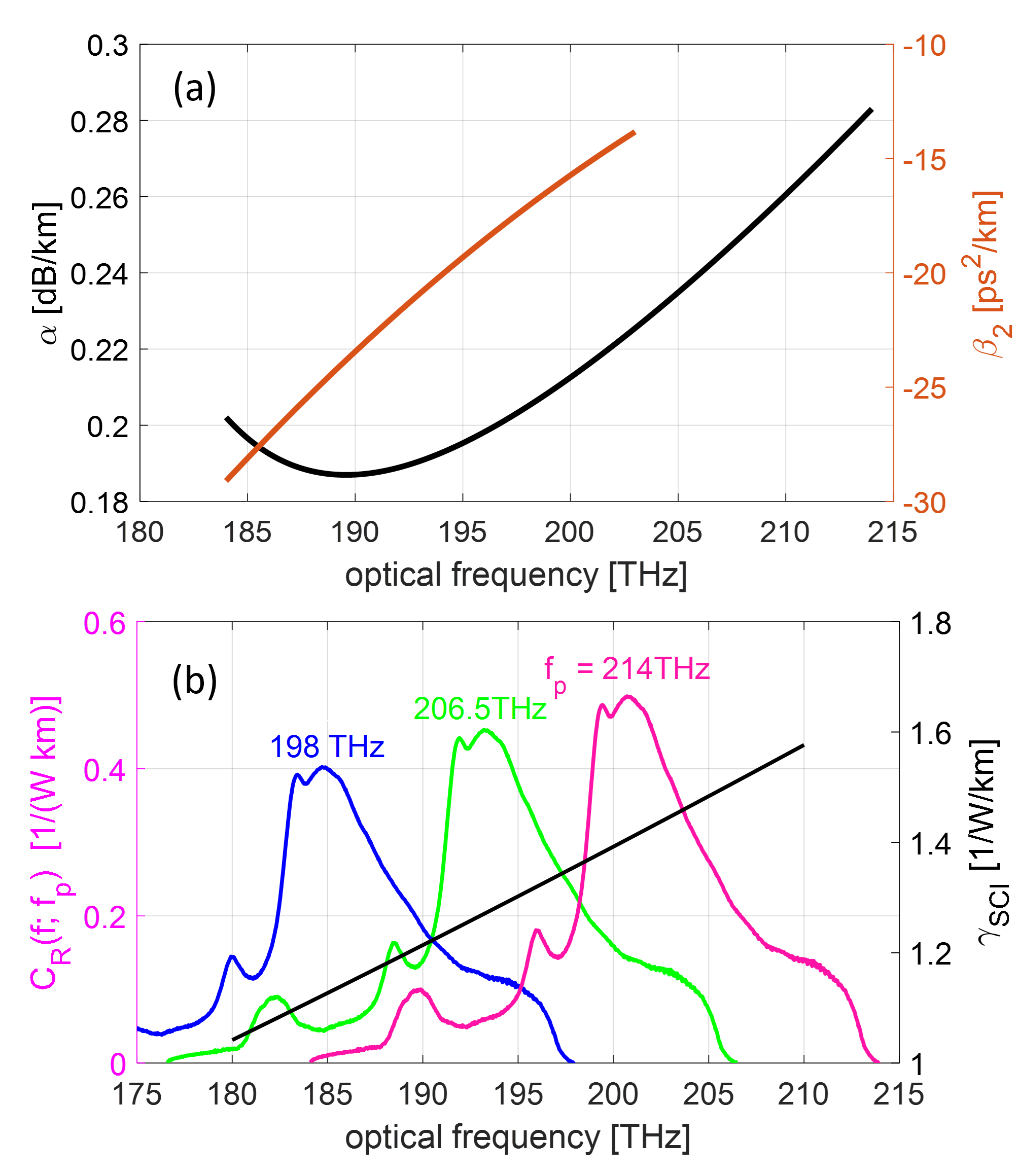}
    \caption{(a) Loss and dispersion; (b) Raman gain spectrum and SCI coefficient $\gamma$.}
    \label{fig:loss_and_dispersion_Raman_gamma}
\end{figure}

\section{PCFM applications and tests}
\label{sec:PCFM_applications}
The polynomial representation of the SPPs in PCFM opens up several avenues of application that have been difficult to deal with otherwise. The problem of accounting for backward-pumped Raman amplification, which required cumbersome and unreliable approximations, is overcome. Low loss fibers, ISRS, forward Raman, but also lumped loss and other critical conditions can be dealt with using these formulas. In the following, we focus on testing PCFM in single-span systems that exhibit the above SPPs behaviors, including lumped loss.

The general methodology of using polynomials has great potential even in situations such as low dispersion or small symbol rates, but this requires using XCI formulas obtained without resorting to stretching the islands  to infinity, and the inclusion of MCI islands too. Thanks to the SPP polynomial approach, this is possible in closed-form too, but we leave these special applications for a forthcoming submission.

To carry out the testing of the PCFM presented here, we first define a test system which will be our reference throughout.

\begin{figure}
    \centering
    \includegraphics[width=0.9\linewidth]{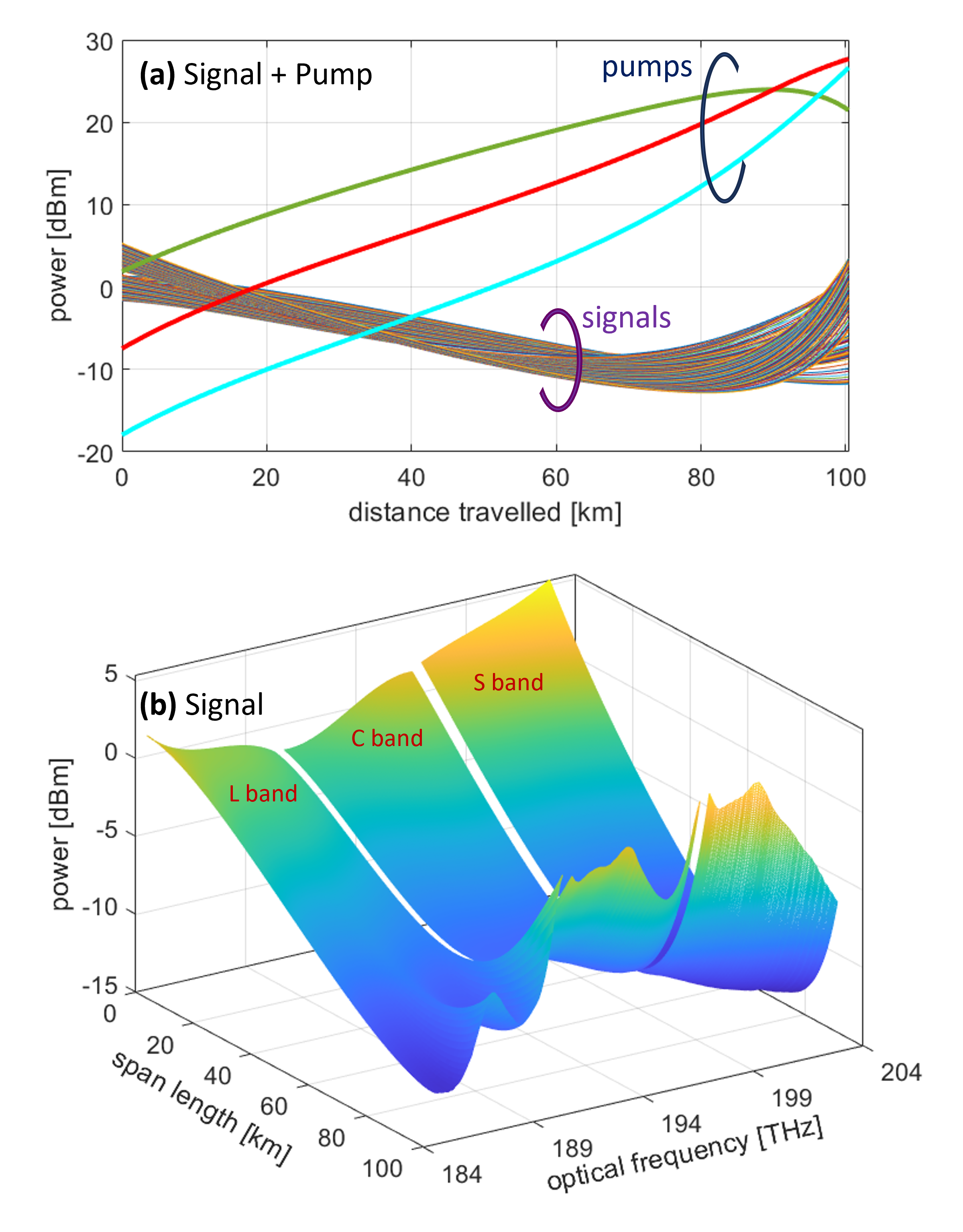}
    \caption{Power evolution along the 100~km fiber. (a): 2D plot of all channels and pumps. (b): 3D plot of all channels.}
    \label{fig:3bands_SPP_100km}
    \vspace{-0.1cm}
\end{figure}

\subsection{Test system description}
\label{sec:system_description}
The system schematic is shown in Fig.~\ref{fig:schematic}. It consists of one span of 100~km, which was the first span in the 10-span system in \cite{2025_JLT_Jiang_SPP}. It was characterized from the experimental set-up used for CFM5 validation in \cite{2025_JLT_Jiang_exp}. Loss and dispersion were measured in the C and L bands and then extrapolated to the S band (Fig.~\ref{fig:loss_and_dispersion_Raman_gamma}(a)) using well-known formulas \cite{1986_JLT_Walker}. The Raman gain spectrum $C_R(f, f_p)$ was experimentally characterized using a pump at $f_p$ = 206.5~THz (the green curve in Fig.~\ref{fig:loss_and_dispersion_Raman_gamma}(b)). It was shifted and scaled as a function of $f$ and $f_p$, according to \cite{2003_JLT_Rottwitt}. The black line in Fig.~\ref{fig:loss_and_dispersion_Raman_gamma}(b) shows the SCI Kerr non-linearity coefficient $\gamma_{\rm SCI}$ of Eq.(\ref{eq:gamma_SCI}) vs.~frequency. Its values and those of $\gamma_{\rm XCI}$ were obtained as indicated next to Eq.~(\ref{eq:gamma_XCI}).

The WDM band boundaries set here were: L-band 184.50 to 190.35; C-band 190.75 to 196.60; S-band 197.00 to 202.85. Each band contained 50 equally-spaced channels, with a symbol rate of 100~GBaud, roll-off 0.1, and spacing 118.75~GHz. The modulation was taken to be Gaussian-shaped. 

Backward Raman pumps were included, consisting of three pumps at 205.1~THz, 21.5~dBm; 211.5~THz, 27.7~dBm; 214.0~THz, 26.6~dBm. Their frequency and power were jointly optimized with the launch power of each channel, to  maximize the overall system throughput, please see Fig.4~(a) of \cite{2025_JLT_Jiang_SPP}. The resulting optimum signal launch power is shown in Fig.~\ref{fig:3bands_SPP_100km}. The figure clearly shows significant Raman amplification, especially in the C and S bands, as well as obvious signs of ISRS, especially in the L-band, where the low-frequency channels initially propagate at zero loss for a few km.

\subsection{Accuracy of the SPP polynomial representation}
\label{sec:SPP_polyfit}
In multiband systems, the computation of the SPPs is typically based on numerically solving the coupled differential Raman equations, which account for both ISRS and Raman amplification \cite{2002_PTL_Perlin}. Conventional solvers, such as the bvp4c algorithm in MATLAB, provide accurate results but they are not optimized for this specific use. We developed a faster SPP estimation algorithm, which we recently reported in \cite{2025_JLT_Jiang_SPP}. The SPPs in Fig.~\ref{fig:3bands_SPP_100km} are computed using such algorithm.

\begin{figure*}
    \centering
    \includegraphics[width=0.85\linewidth]{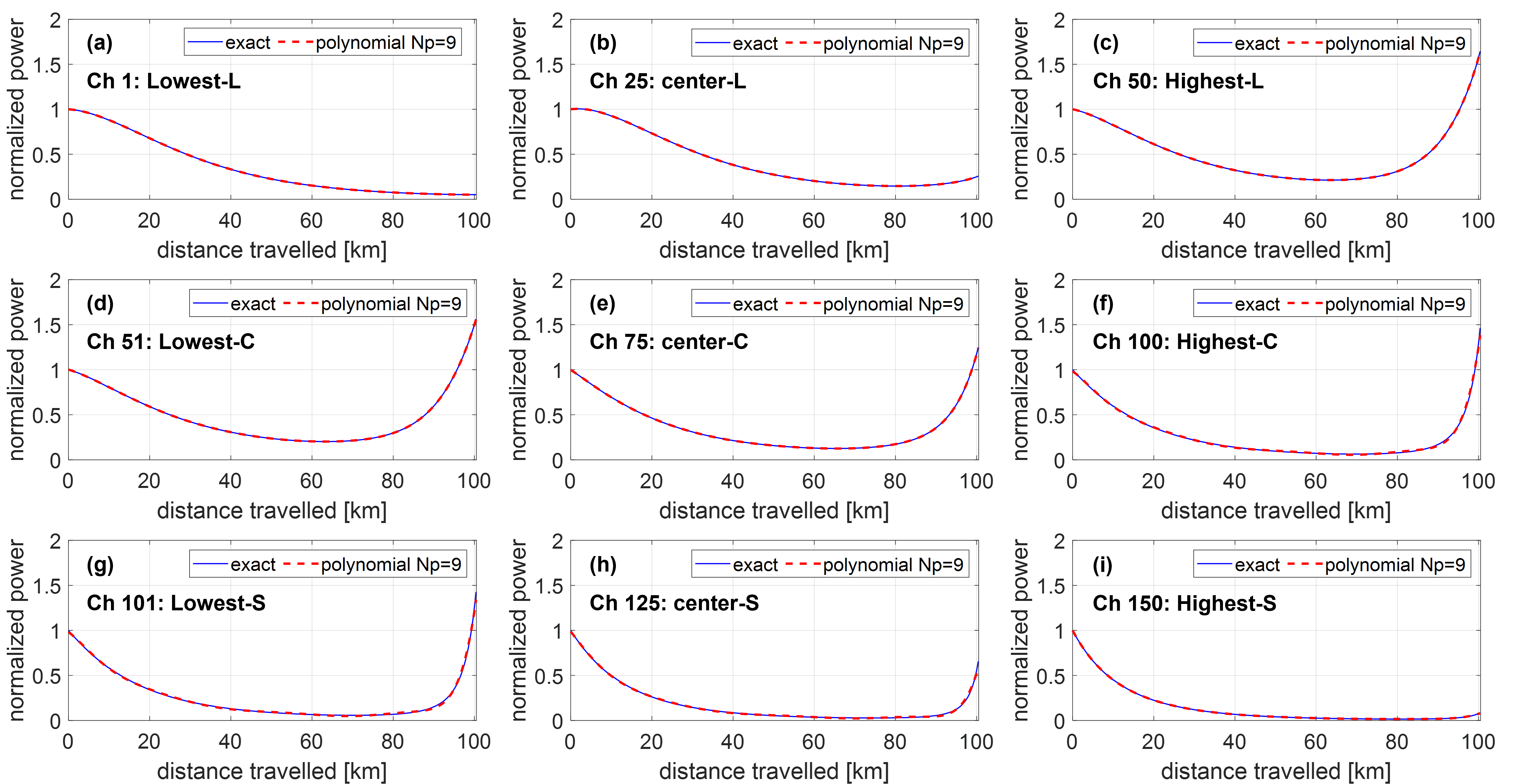}
     \vspace{0.2cm}
    \caption{Polynomial representation of normalized SPPs along the 100~km fiber in L, C and S bands: blue solid curves are the exact SPPs, and red dashed curves are the polynomial fit with $N_p$ = 9.}
    \label{fig:SPP_polyfit_Np9_100km}
\end{figure*} 

Starting from the SPPs, a 9th-degree polynomial is then used to bestfit the SPP of each channel. Fig.~\ref{fig:SPP_polyfit_Np9_100km} displays the polynomial fit for three channels in each band. 

Note that the L-band channels experience very low loss, or no loss, at the start, while the S-band channels experience high loss at the start. This is only partially due to the fiber having lower intrinsic loss in the L-band than in the S-band. Much of the behavior is due to the S-band channels losing power to the benefit of the L-band channels, because of ISRS. C-band and S-band channels also experience substantial gain at the end of the fiber due to backward Raman amplification.

The $N_p$ = 9 polynomial fit of Fig.~\ref{fig:SPP_polyfit_Np9_100km} appears excellent across all the different SPPs. To provide an idea of what type of error can otherwise be incurred in SPP fitting when using lower degree polynomials, we focus on one of the most challenging channels of the system. It is channel 100 of Fig.~\ref{fig:SPP_polyfit_Np9_100km}(f), which exhibits a very steep power increase towards the end of the span. Its polynomial fitting with $N_p$=1,~3,~5 and 7 is displayed in Fig.~\ref{fig:SPP_polyfit_Nch100}. Expectedly, degrees 1, 3 do not look good. As for 5, it still does not appear fully satisfactory. 

However, when the actual error on NLI estimation is computed, it turns out that rather accurate results can be obtained even for relatively low values of $N_p$.  This seem to indicate that some deviation of the polynomial fitting from the SPP, such as visible for $N_p$=5, may be `averaged out' during integration and eventually have  limited impact on NLI assessment. We will come back to this in detail in the next section.

\subsection{Accuracy of NLI estimation}
\label{sec:err_vs_Np}

\begin{figure}
    \centering
    \includegraphics[width=1.0\linewidth]{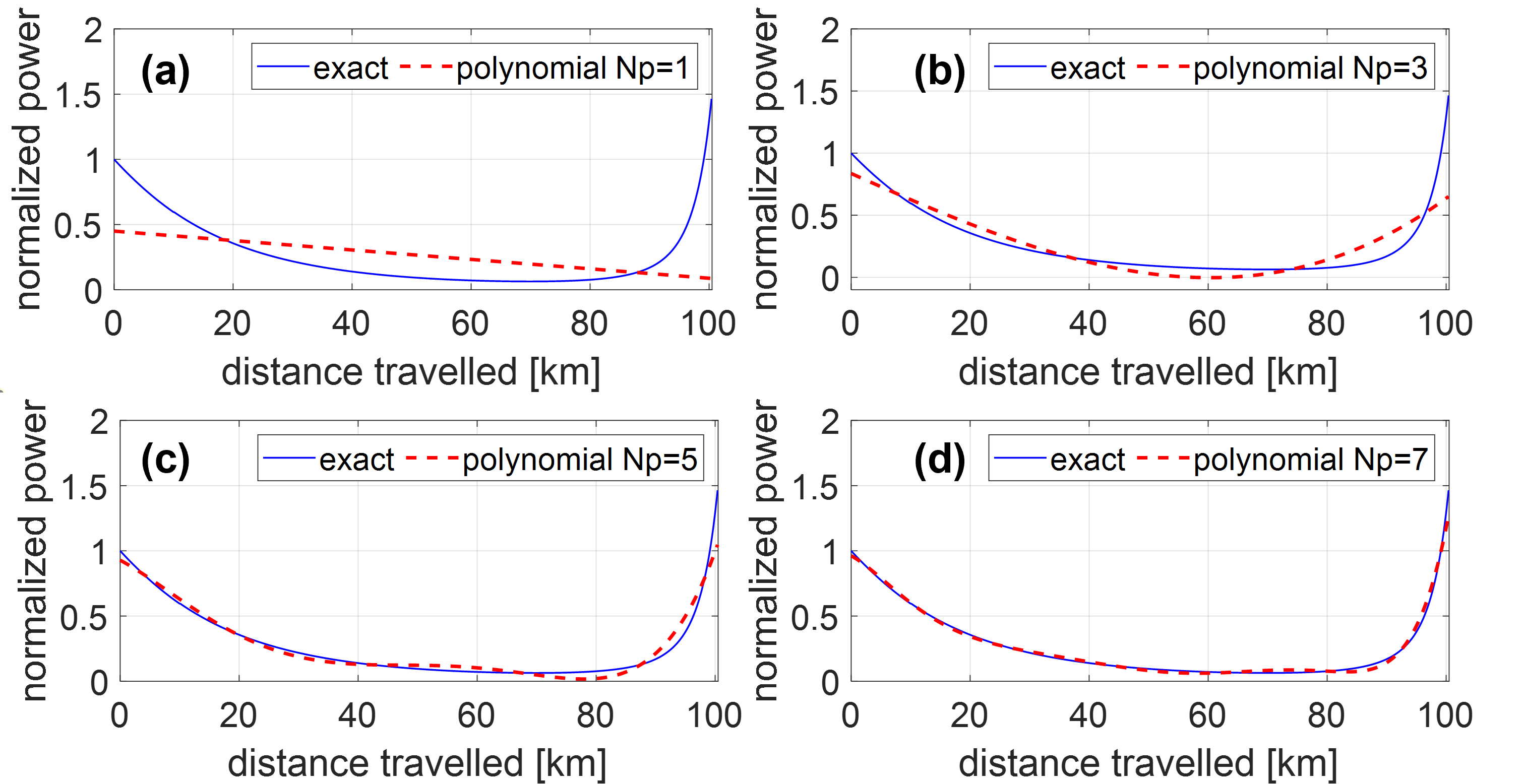}
     \vspace{0.1cm}
    \caption{Polynomial representation of the normalized SPP along the 100~km fiber in the 100-th (highest-frequency C-band) channel: the blue solid curves are the exact SPPs, and the red dashed curves are the polynomial fit with (a) $N_p$ = 1,  (b) $N_p$ = 3,  (c) $N_p$ = 5,  (d) $N_p$ = 7.}
    \label{fig:SPP_polyfit_Nch100}
\end{figure}

After SPP polynomial fitting, we use the PCFM to compute a key NLI indicator for all 150 channels in the system. We focus on $\text{GSNR}_{\text{NLI}}$, defined as: 
\begin{equation}
    \text{GSNR}_{\text{NLI}} = \frac{P_{\text{ch}}}{P_{\text{NLI}}}
    \label{eq:GSNR_NLI}
\end{equation}
We then compare the result of $\text{GSNR}_{\text{NLI}}$ obtained through the PCFM with that of the numerically-integrated full EGN model \cite{2014_OE_Carena}, which is our benchmark. We call $\Delta\text{GSNR}_{\text{NLI}}$ the error, in dB scale:
\begin{equation}
    \Delta\text{GSNR}_{\text{NLI}}  = \text{GSNR}_{\text{NLI,PCFM}} - \text{GSNR}_{\text{NLI,EGN}}
    \label{eq:delta_GSNR_NLI}
\end{equation}

 \begin{figure}
    \centering
    \includegraphics[width=0.9\linewidth]{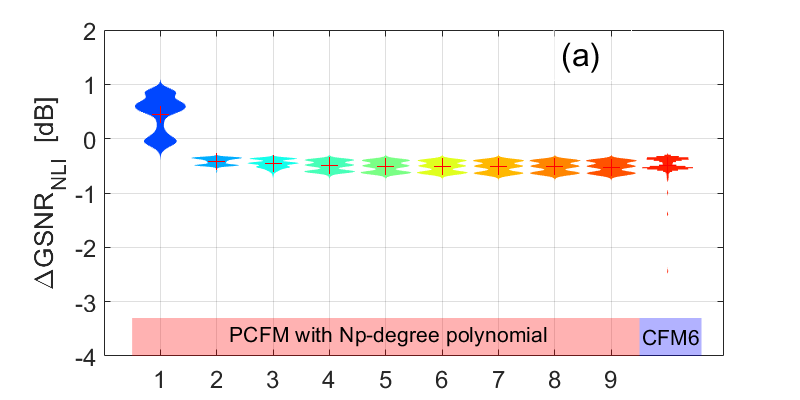}
    \includegraphics[width=0.9\linewidth]{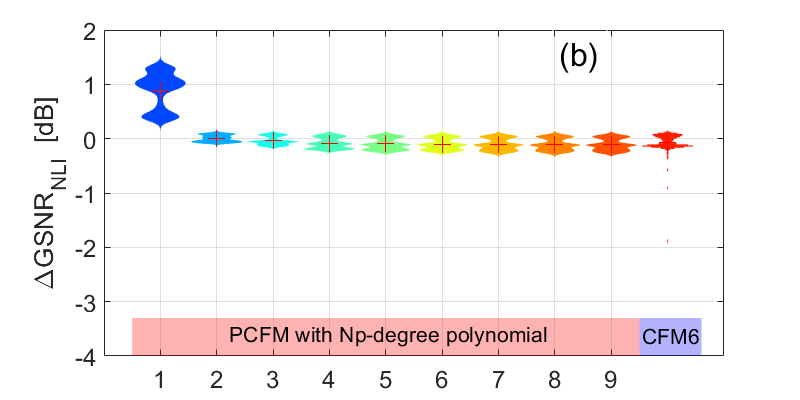}
    \caption{NLI estimation error of PCFM vs. the GN-model, in the 100~km C+L+S system. \textbf{(a)}: violin plot of $\Delta\text{GSNR}_{\text{NLI}}$ calculated using Eq.~(\ref{eq:delta_GSNR_NLI}), where PCFM uses $N_p$ = 1 to 9; \textbf{(b)} violin plot of $\Delta\text{GSNR}_{\text{NLI}}$ with the machine-learning correction $\rho$ introduced in Eq.~(14) of \cite{2020_JLT_Zefreh}. In both plot the $\Delta\text{GSNR}_{\text{NLI}}$ of CFM6 vs. the GN-model is also shown.}
    \label{fig:violin_plot_100km}
\end{figure}

Before looking at the result, we would like to remind the Readers that an error of $\delta$ dB in $\rm GSNR_{\rm NLI}$ estimation causes an error in total $\rm GSNR$ of only $\delta/3$ dB, if operating at the optimum launch power. This is a well-known result from the early days of NLI modeling \cite{2011_OFC_Bosco}, \cite{2011_OE_Grellier}, which has helped make the use of CFMs more viable over the years, by substantially quelling the error coming from the approximations needed to achieve closed-forms. The factor 1/3 should be kept in mind while looking at any NLI assessment accuracy, and we are going to call it the `1/3 rule' henceforth.

 Fig.~\ref{fig:violin_plot_100km}(a) shows the violin plot of $\Delta\text{GSNR}_{\text{NLI}}$ across all channels, where different polynomial degrees are used to fit the SPPs. As $N_p$ increases, the error distribution stabilizes already for $N_{p}$=~4. This is a somewhat unexpected result given that the plots of Fig.~\ref{fig:SPP_polyfit_Nch100} show significant SPP fitting error still for $N_{p}$=~5. We conjecture that the repeated integration in the core integrals $K_x^{(n_s)} (f_{\text{CUT}})$ averages out the fitting errors in the SPPs. Notwithstanding, the violin plots of Fig.~\ref{fig:violin_plot_100km}(a) suggest that, even in complex systems like the one being probed, relatively low values of $N_{p}$ are potentially usable to get fast preliminary NLI estimation, if speed is critical. The final results could then be validated using higher degrees, such as $N_{p}$=~9.

The violin plots for $N_{p}\ge 4$ have a very small standard deviation of 0.1~dB, an excellent result. However, the picture shows a systematic bias of approximately $-0.5$~dB, meaning that the PCFM overestimates NLI by 0.5~dB, leading to a lower $\rm GSNR_{\rm NLI}$ than the EGN benchmark. This systematic overestimation arises from the two main approximations described in detail in the previous sections, namely: assuming that the NLI spectrum over the CUT is flat at the value that it assumes at the center frequency of the CUT (whereas it tapers off somewhat); approximating the lozenge-shaped integration islands with bigger shapes. On the other hand, the 1/3 rule means that the bias on the \textit{total} GSNR at optimum launch power would be only about $-0.2$~dB. Note also that, since negative, it would be conservative.

 Nonetheless, it would be nice to remove such bias. In previous CFM versions we confronted a similar problem and resorted to an analytical machine-trained correction, which we called the `machine-learning coefficient' $\rho$, introduced in Eq.~(14) of \cite{2020_JLT_Zefreh}. We decided to apply such coefficient to the PCFM too, with no change vs. \cite{2020_JLT_Zefreh}. As the the violin diagrams of Fig.~\ref{fig:violin_plot_100km}(b) show, the bias disappears almost completely. We are currently re-training $\rho$ specifically for the PCFM so that it is possible that even better results, for both residual bias and standard deviation, could be achieved. 

The plots of Fig.~\ref{fig:violin_plot_100km} also show the violin diagram for CFM6. The result for most channels is comparable to those of the PCFM but there are several outliers whose error is as large as -2.5 dB. This is due to a problem intrinsic to the analytical derivation of CFM6 where a series expansion followed by integration produces a denominator that can go to zero. When calculating NLI values near these points, we conjecture that loss of machine precision occurs, leading to errors. While such singularities are removable by means of ad-hoc workarounds, and the outliers can be brought within the violin main body, there remains the need to check for their occurrence during the calculations and then apply the workarounds, a procedure which is cumbersome. This is one of the reasons that compelled us to pursue a new approach, resulting in the PCFM, which is free from any singularity. 

There were other reasons too, as mentioned before, such as the difficulty or impossibility of upgrading CFM6 to situations such as low loss, lumped loss, short spans, multi-subcarrier multiplexing and ultra-low dispersion, where PCFM provides a solution or promises to make one possible. One more important reason had to do with modeling intra-span NLI coherence. This aspect is clarified in the the following.

\begin{figure}
    \centering
\includegraphics[width=0.85\linewidth]{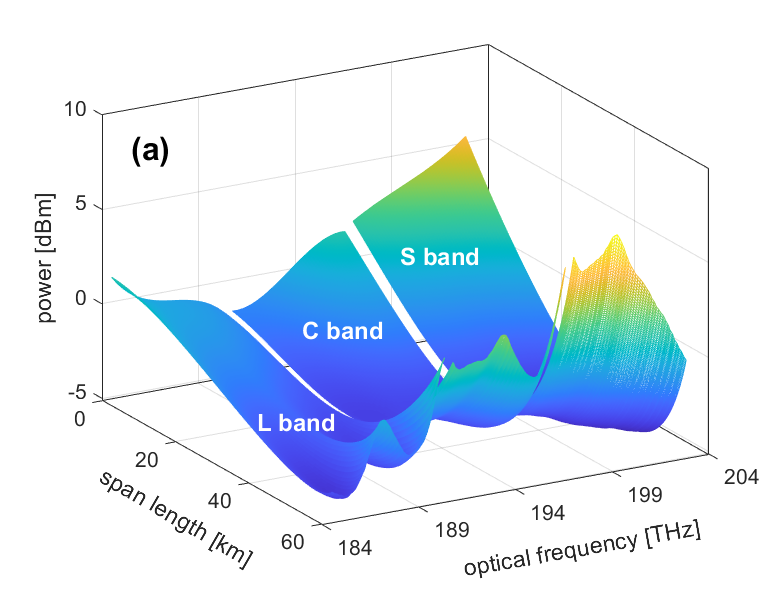}
\includegraphics[width=0.85\linewidth]{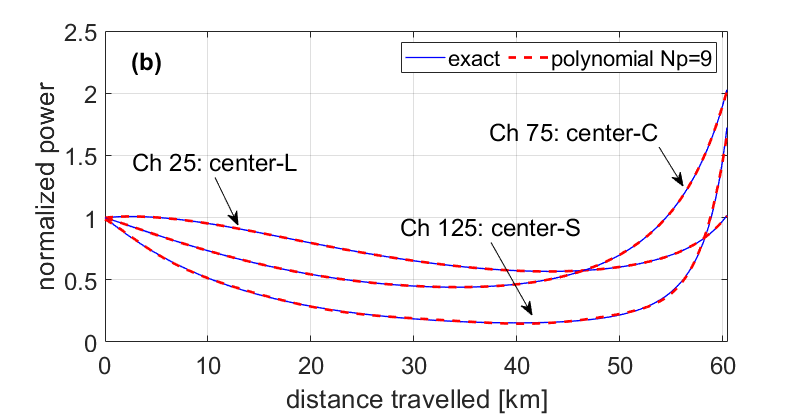}
\includegraphics[width=0.85\linewidth]{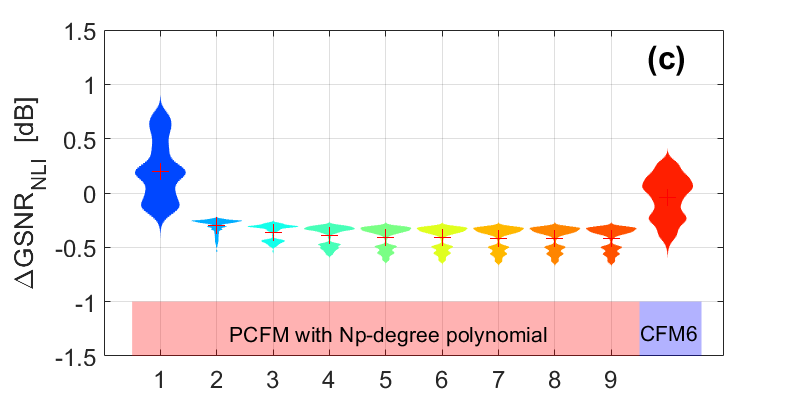}
\includegraphics[width=0.85\linewidth]{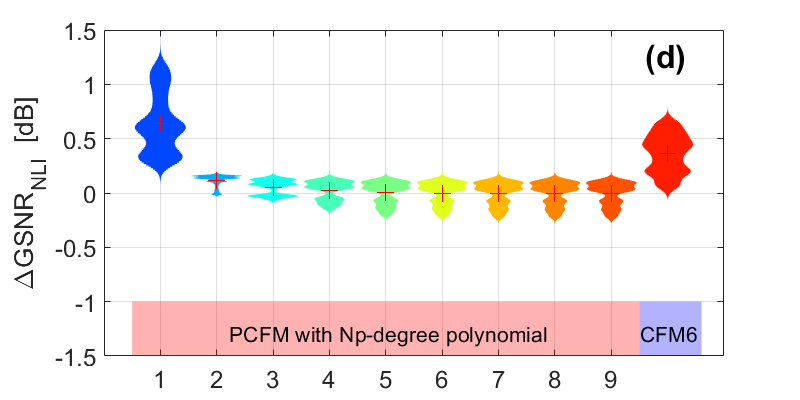}
    \caption{60~km fiber system example. \textbf{(a):} 3D plot of all channel SPPs. \textbf{(b):} 9th-degree polynomial representations of the center-channel SPPs in each band. \textbf{(c):} violin plot of PCFM and CFM6 results. \textbf{(d):} violin plots with machine-learning correction incorporated in CFMs.}
    \label{fig:3bands_60km}
\end{figure}

\begin{figure}
    \centering
    \includegraphics[width=0.85\linewidth]{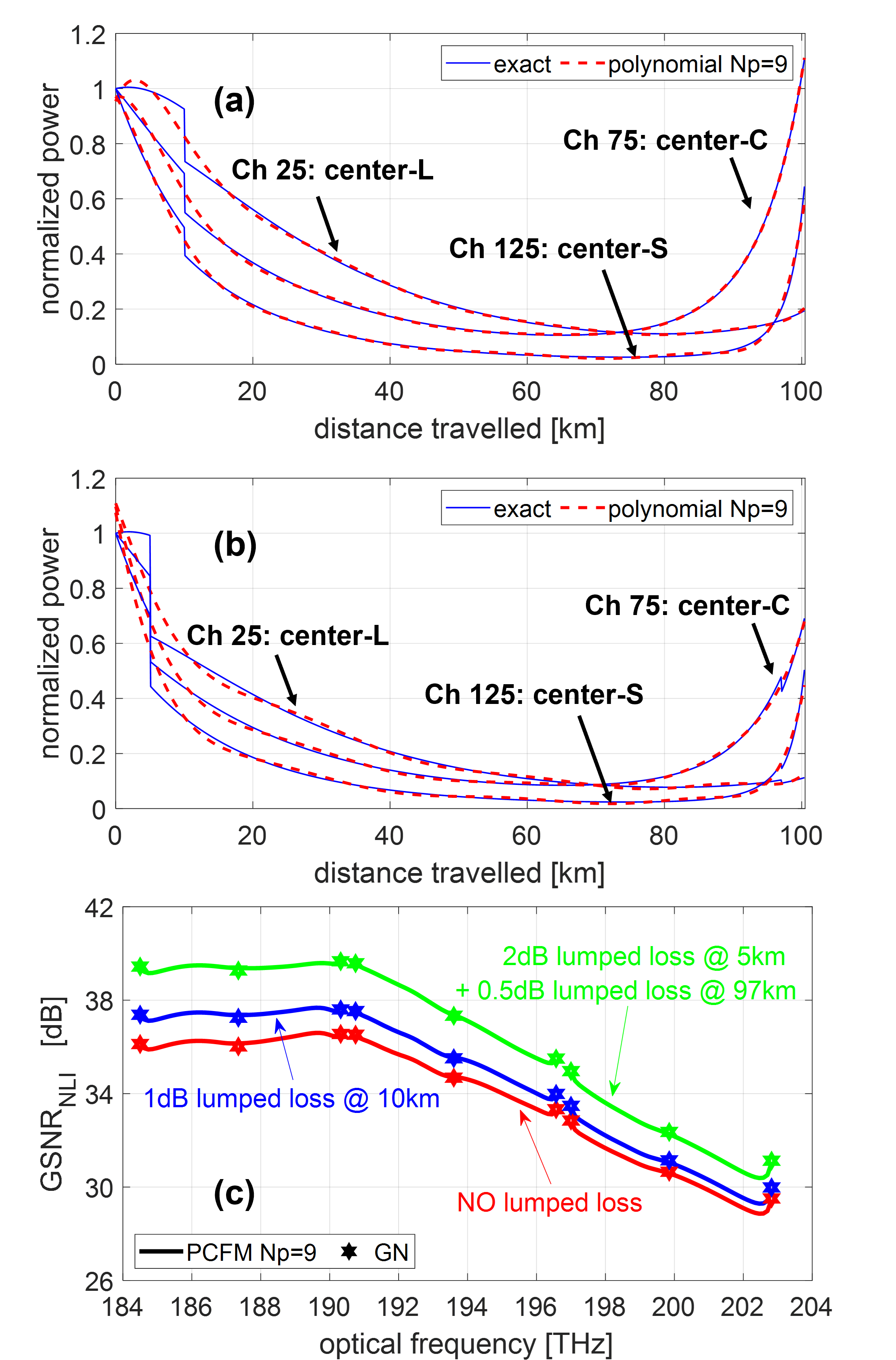}
    \vspace{5mm}
    \caption{Results with lumped loss. \textbf{(a):} 9th-degree polynomial fitting of the center-channel SPPs in each band, with 1~dB lumped loss at 10~km. \textbf{(b):} 9th-degree polynomial fitting of the center-channel SPPs in each band, with added 2~dB lumped loss at 5~km and 0.5~dB lumped loss at 97~km. \textbf{(c):} $\text{GSNR}_{\text{NLI}}$ calculated by the PCFM for all channels. The markers are the numerically-integrated GN-model benchmark results.}
    \label{fig:results_loss}
\end{figure}

We explore a new test scenario where we reduce the fiber length to 60~km. All SPPs are shown in Fig.~\ref{fig:3bands_60km}~(a). Some are shown again in detail, together with their 9-th degree polynomial fit, in Fig.~\ref{fig:3bands_60km}~(b). From the plots, it can be seen that with such shorter fiber length, in the presence of ISRS and Raman amplification, some SPPs experience almost ideal distributed amplification. For instance, channel 25 in the L-band sees a maximum loss of less than 3~dB, and its overall profile along the fiber is remarkably flat. 

The violin plot of $\Delta\text{GSNR}_{\text{NLI}}$ is depicted in Fig.~\ref{fig:3bands_60km}~(c). The distribution gets stable at $N_p$ = 5, again with a quite contained standard deviation of about 0.1~dB. As in the 100~km example, a $-0.5$~dB bias is present. Once again, applying the machine-learning coefficient $\rho$ from \cite{2020_JLT_Zefreh} the bias is eliminated (Fig.~\ref{fig:3bands_60km}~(d)).

What is different between the violin plots of Fig.~\ref{fig:3bands_60km} and Fig.~\ref{fig:violin_plot_100km} is that CFM6 has a much wider distribution than PCFM, which remains such also after applying the machine-learning coefficient. Note that this greater error is not due to the removable singularity problem of CFM6 previously mentioned. Here it is a fundamental shortcoming, which we explain in the following. 

CFM6 resorts to breaking up the span into two sub-spans, approximately in the middle of the span. It then separately calculates NLI for both sub-spans. The two contributions are incoherently summed at the end of the span. This is done to allow CFM6 to model the NLI produced by the power growth at the end of the fiber, because of backward-pumped Raman amplification.  As Fig.~\ref{fig:violin_plot_100km} shows, the approach works well in the 100~km case, where mid-span loss is substantial (8-10 dB, see Fig.~\ref{fig:SPP_polyfit_Np9_100km}). When this is not the case, such as in the 60~km system, where mid-span loss is as low as 3~dB, a quite substantial error is incurred by CFM6, as visible in Fig.~\ref{fig:3bands_60km}~(c) and (d), because of the neglected coherence between the NLI contributions of the two sub-spans. PCFM has no such problem, since the span is dealt with in its entirety, faithfully accounting for intra-span NLI coherence.

\subsection{SPP polynomial representation and lumped loss}
\label{sec:lumped_loss}

Polynomials are clearly effective in accurately describing SPPs subject to `smooth' influences, such as ISRS or Raman gain. It is not immediately evident that they could be equally effective when abrupt or `lumped' perturbations are experienced, such as the case of lumped loss. 

To assess this capability, we revisit the 100~km configuration and introduce lumped losses of varying magnitudes at different locations along the fiber. Fig.~\ref{fig:results_loss}~(a) displays the SPPs of the center channels of each of the three bands, with a 1~dB lumped loss introduced at 10~km, such as due to a bad splice or a lossy connector.

Even though an intrinsically continuous curve cannot perfectly fit a discontinuous profile, for $N_{p}=9$ it appears that the polynomial fit does a reasonable job of closely interpolating the SPP. The polynomial curve then goes through repeated integration and it is conceivable that SPP fitting inaccuracy can be averaged out. The resulting GSNR prediction by the PCFM indeed aligns closely with the GN-model markers, as shown by the blue curve in Fig.~\ref{fig:results_loss}~(c), with no apparent loss of accuracy with respect to the case of no lumped loss.

A more extreme test includes \textit{two} lumped losses along the span: 2~dB at 5~km and and 0.5~dB at 97~km. Both losses have a significant impact on the SPPs, the first directly and the second because it affects both the signal and the Raman pumps power. Here too the SPPs are not perfectly fitted by the polynomials. However, the interpolation they provide results again in a final GSNR prediction by PCFM (green curve in Fig.~\ref{fig:results_loss}~(c)) which appears as accurate as the no-loss case.

Overall, considering the complexity of the SPP shapes of Fig.~\ref{fig:results_loss}~(a) and (b), characterized by substantial ISRS, strong backward-pumped Raman amplification and multiple lumped loss, resulting in profiles that are very far from the simple decreasing exponential of a conventional fiber, we consider the accuracy displayed by the NLI prediction of Fig.~\ref{fig:results_loss}~(c) as remarkable. Note that the plot displays $\rm GSNR_{NLI}$, so there is no 1/3-rule at play here to attenuate the error.

\subsection{Non-Gaussian formats and inter-span coherence}
\label{sec:multi_span_and_EGN}

So far we have focused on the discussion of the PCFM performance in the context of Gaussian-shaped modulation.

As debated previously, the PCFM, like CFM6 and essentially all CFMs in the literature, is derived from the GN-model, which is most accurate when Gaussian-shaped modulation is used. One could use PCFM `as is' to model QAM systems, but there would be some error. Such error is relatively modest in links with high dispersion, long spans and high symbol rates, but it is present and it is the reason why the EGN-model was developed \cite{2013_OE_Dar}, \cite{2014_OE_Carena}.

To achieve accurate format-dependent, EGN-like prediction with a GN-model derived CFM, we developed in the past suitable corrections which account for the effect of different formats, as well as the coherence of NLI among multiple spans (inter-span coherence), please see \cite{2020_JLT_Zefreh} and the experimental validations \cite{2024_ECOC_Jiang}, \cite{2025_JLT_Jiang_exp}. We are also in the process of developing and training an improved set of corrections, in the quest for greater accuracy, which we leave for a forthcoming paper. Due also to space constraints, we do not pursue this topic further here.

\section{Downloadable software and developments}
\label{sec:software}
We have implemented in software the SCI+XCI PCFM model described in this paper, with SPP fitting polynomial degree $N_{p}\!=\!9$. It includes the corrections mentioned above for multi-span and multi-format (EGN-model) support. 

The software is quite flexible and allows to estimate the $ \rm GSNR_{NLI}$ and total $ \rm GSNR$  of rather general WDM UWB systems, where each span can be different from any other. It also handles forward and backward-pumped Raman amplification, as well as ISRS. ASE is computed for lumped amplification as well as Raman. Fiber data is fully customizable, span by span. The WDM comb does not need to be uniform: channels can have all different: spacing, symbol rate, modulation format and launch power. The launch powers can change span by span.

Currently, the main file and parameter file are in editable Matlab code, whereas the model is provided as executable files for Windows and Linux \cite{2025_zenodo_PCFM}. It is available for download under the Creative Commons 4.0 License. Editing the main file and parameter file provides full system customization flexibility.

We will be upgrading it to provide more pre-programmed examples and to add further features. Future releases will include Double Rayleigh Backscattering, XCI without the island-stretching approximation, as well as MCI with rectangular islands, to support low dispersion and multi-subcarrier systems.

\section{Conclusion}

We have presented a closed-form GN/EGN model, called Polynomial-CFM or PCFM, based on a new analytical approach consisting of expressing the spatial power profile, or SPP, of each channel as a polynomial. Once this is done, remarkably, closed-form analytical solutions of the core integrals of the GN-model can be found with few approximations. This leads to improved CFM reliability and generality. The polynomial SPP representation allows to model quite general system conditions: we have shown results and examples of the accuracy and generality of the PCFM, for UWB systems with ISRS and Raman amplification, as well as lumped loss along the span. 

The polynomial SPP approach promises to allow tackling even more general system configurations, that were difficult to deal with by previous CFMs, such as low symbol rates, multi-subcarrier transmission, very low dispersion and others.

We provide a software implementation of the model described in this paper, available for download and fully customizable.



\begin{thebibliography}{00}

\bibitem{2024_ECOC_Puttnam}
B.~J. Puttnam \emph{et al.}, ``339.1~Tb/s OESCLU-band transmission over 100~km SMF,'' in \emph{Proc. Eur. Conf. Opt. Commun. (ECOC)}, Frankfurt, Germany, 2024, paper M2B.2.

\bibitem{2024_PTL_Jiang}
Y.~Jiang, J.~Sarkis, A.~Nespola, F.~Forghieri, S.~Piciaccia, and A.~Tanzi, \emph{et al.}, ``Optimization of long-haul C{+}L{+}S systems by means of a closed-form EGN model,'' \emph{IEEE Photon. Technol. Lett.}, vol.~36, no.~18, pp.~1129--1132, 2024.

\bibitem{2024_ECOC_Kim}
I.~Kim \emph{et al.}, ``Band-wise bidirectional S{+}C{+}L transmission in hybrid Raman--EDFA link,'' in \emph{Proc. Eur. Conf. Opt. Commun. (ECOC)}, Frankfurt, Germany, 2024, paper M3B.1.

\bibitem{2025_JLT_Puttnam}
B.~J. Puttnam \emph{et al.}, ``High data-rate OESCLU-band transmission,'' \emph{J. Lightw. Technol.}, early access, pp.~1--9, 2025, doi: 10.1109/JLT.2025.3543448.

\bibitem{2025_OFC_Shimizu}
S.~Shimizu \emph{et al.}, ``27-THz ISRS-supported transmission over 1040~km in S{+}C{+}L{+}U and extreme longer-wavelength band,'' in \emph{Proc. Opt. Fiber Commun. Conf. (OFC)}, San Francisco, CA, USA, 2025, paper Th4A.2.

\bibitem{2025_OFC_Sarkis}
J.~Sarkis, Y.~Jiang, S.~Piciaccia, F.~Forghieri, and P.~Poggiolini, ``Signal and Raman pump launch power optimization in a C{+}L{+}S{+}E system using fast power profile estimation,'' in \emph{Proc. Opt. Fiber Commun. Conf. (OFC)}, San Francisco, CA, USA, 2025, paper Tu3K.3.

\bibitem{2023_JLT_Buglia}
H.~Buglia \emph{et al.}, ``A closed-form expression for the Gaussian noise model in the presence of inter-channel stimulated Raman scattering extended for arbitrary loss and fibre length,'' \emph{J. Lightw. Technol.}, vol.~41, no.~11, pp.~3577--3586, 2023.

\bibitem{2024_JLT_Buglia}
H.~Buglia, M.~Jarmolovicius, L.~Galdino, R.~I. Killey, and P.~Bayvel, ``A closed-form expression for the Gaussian noise model in the presence of Raman amplification,'' \emph{J. Lightw. Technol.}, vol.~42, no.~2, pp.~636--648, 2024.

\bibitem{2022_ECOC_Poggiolini}
P.~Poggiolini and M.~Ranjbar-Zefreh, ``Closed-form expressions of the nonlinear interference for UWB systems,'' in \emph{Proc. Eur. Conf. Opt. Commun. (ECOC)}, Basel, Switzerland, 2022, paper Tu1D.1. 

\bibitem{2024_ECOC_Jiang}
Y.~Jiang, A.~Nespola, S.~Straullu, A.~Tanzi, S.~Piciaccia, F.~Forghieri, D.~Pilori, and P.~Poggiolini, ``Closed-form EGN model with comprehensive Raman support,'' in \emph{Proc. Eur. Conf. Opt. Commun. (ECOC)}, Frankfurt, Germany, 2024, paper W1B.1.

\bibitem{2022_JLT_DAmico}
A.~D'Amico \emph{et al.}, ``Scalable and disaggregated GGN approximation applied to a C{+}L{+}S optical network,'' \emph{J. Lightw. Technol.}, vol.~40, no.~11, pp.~3499--3511, 2022.

\bibitem{2022_OE_Shevchenko}
N.~Shevchenko, S.~Nallaperuma, and S.~Savory, ``Maximizing the information throughput of ultra-wideband fiber-optic communication systems,'' \emph{Opt. Express}, vol.~30, no.~11, pp.~19320--19331, 2022.

\bibitem{2023_JLT_Lasagni}
C.~Lasagni \emph{et al.}, ``A generalized Raman scattering model for real-time SNR estimation of multi-band systems,'' \emph{J. Lightw. Technol.}, vol.~41, no.~11, pp.~3407--3416, 2023.

\bibitem{2012_JLT_Poggiolini}
P.~Poggiolini, ``The GN model of non-linear propagation in uncompensated coherent optical systems,'' \emph{J. Lightw. Technol.}, vol.~30, no.~24, pp.~3857--3879, 2012, doi: 10.1109/JLT.2012.2217729. 

\bibitem{2014_OE_Carena}
A.~Carena, G.~Bosco, V.~Curri, Y.~Jiang, P.~Poggiolini, and F.~Forghieri, ``EGN model of non-linear fiber propagation,'' \emph{Opt. Express}, vol.~22, no.~13, pp.~16335--16362, Jun. 2014, doi: 10.1364/OE.22.016335.

\bibitem{2025_OFC_Poggiolini}
P.~Poggiolini and Y.~Jiang, ``Recent advances in real-time models for UWB transmission systems,'' in \emph{Proc. Opt. Fiber Commun. Conf. (OFC)}, San Francisco, CA, USA, 2025, paper Tu3K.2.

\bibitem{2020_JLT_Zefreh}
M.~R. Zefreh, F.~Forghieri, S.~Piciaccia, and P.~Poggiolini, ``Accurate closed-form real-time EGN model formula leveraging machine-learning over 8500 thoroughly randomized full C-band systems,'' \emph{J. Lightw. Technol.}, vol.~38, no.~18, pp.~4987--4999, 2020.

\bibitem{2010_OE_Santagiustina}
M.~Santagiustina, C.~G. Someda, G.~Vadal\`a, S.~Combri\'e, and A.~De~Rossi, ``Theory of slow light enhanced four-wave mixing in photonic crystal waveguides,'' \emph{Opt. Express}, vol.~18, no.~20, pp.~21024--21029, Sept. 2010, doi: 10.1364/OE.18.021024.

\bibitem{2003_JLT_Rottwitt}
K.~Rottwitt, J.~Bromage, A.~J. Stentz, L.~Leng, M.~E. Lines, and H.~Smith, ``Scaling of the Raman gain coefficient: applications to germanosilicate fibers,'' \emph{J. Lightw. Technol.}, vol.~21, no.~7, pp.~1652--1662, Jul. 2003, doi: 10.1109/JLT.2003.814386.

\bibitem{2020_chapter_Bononi}
A.~Bononi, R.~Dar, M.~Secondini, P.~Serena, and P.~Poggiolini, ``Fiber nonlinearity and optical system performance,'' in \emph{Springer Handbook of Optical Networks}, B.~Mukherjee, I.~Tomkos, M.~Tornatore, P.~Winzer, and Y.~Zhao, Eds. Cham, Switzerland: Springer, 2020, ch.~9. 

\bibitem{1977_BellTech_Marcuse}
D.~Marcuse, ``Loss analysis of single-mode fiber splices,'' \emph{Bell Syst. Tech. J.}, vol.~56, no.~5, pp.~703--718, May--Jun. 1977, doi: 10.1002/j.1538-7305.1977.tb00534.x.

\bibitem{2025_TechRxiv_Poggiolini}
P.~Poggiolini, ``P-CFMs: polynomial power-profile closed-form models related to the GN-model,'' TechRxiv, 2025, doi: 10.36227/techrxiv.174803120.05191968/v2.

\bibitem{1986_JLT_Walker}
S.~Walker, ``Rapid modeling and estimation of total spectral loss in optical fibers,'' \emph{J. Lightw. Technol.}, vol.~4, no.~8, pp.~1125--1131, 1986.

\bibitem{2002_PTL_Perlin}
V.~E. Perlin and H.~G. Winful, ``Optimizing the noise performance of broad-band WDM systems with distributed Raman amplification,'' \emph{IEEE Photon. Technol. Lett.}, vol.~14, no.~8, pp.~1199--1201, 2002.

\bibitem{2011_OFC_Bosco}
G.~Bosco, A.~Carena, R.~Cigliutti, V.~Curri, P.~Poggiolini, and F.~Forghieri, ``Performance prediction for WDM PM-QPSK transmission over uncompensated links,'' in \emph{Proc. Opt. Fiber Commun. Conf. (OFC)}, Mar. 2011, paper OThO7.

\bibitem{2011_OE_Grellier}
E.~Grellier and A.~Bononi, ``Quality parameter for coherent transmissions with Gaussian-distributed nonlinear noise,'' \emph{Opt. Express}, vol.~19, no.~13, pp.~12781--12788, Jun. 2011.

\bibitem{2013_OE_Dar}
R.~Dar, M.~Feder, A.~Mecozzi, and M.~Shtaif, ``Properties of nonlinear noise in long, dispersion-uncompensated fiber links,'' \emph{Opt. Express}, vol.~21, no.~22, pp.~25685--25699, Nov. 2013, doi: 10.1364/OE.21.025685.

\bibitem{2025_JLT_Jiang_exp}
Y.~Jiang, A.~Nespola, S.~Straullu, A.~Tanzi, S.~Piciaccia, F.~Forghieri, and P.~Poggiolini, ``Experimental test of a closed-form EGN model over C{+}L bands,'' \emph{J. Lightw. Technol.}, vol.~43, no.~2, pp.~439--449, 2025, doi: 10.1109/JLT.2024.3455752.

\bibitem{2025_JLT_Jiang_SPP}
Y.~Jiang, J.~Sarkis, S.~Piciaccia, F.~Forghieri, and P.~Poggiolini, ``Signal and backward Raman pump power optimization in multi-band systems using fast power profile estimation,'' \emph{J. Lightw. Technol.}, vol.~43, no.~17, pp.~8140--8149, Sept. 2025, doi: 10.1109/JLT.2025.3585684.

\bibitem{2025_zenodo_PCFM}  Y.~Jiang, Yifeng Gao, Tianchun Zhu, P.~Poggiolini. Optical link performance calculator based on the Polynomial Closed-Form GN/EGN model (PCFM). Zenodo. https://zenodo.org/records/16967985

\end{thebibliography}
\end{document}